\documentclass[aps,prb,twocolumn, superscriptaddress]
{revtex4-1}

\usepackage{amsmath} 
\usepackage{amsthm} 
\usepackage{graphicx} 
\usepackage{float}

\begin{document}
\setlength{\parskip}{0pt}

\title{Surface impedance and optimum surface resistance of a superconductor with imperfect surface}

\author{Alex Gurevich}
\email[]{gurevich@odu.edu}
\affiliation{Department of Physics and Center for Accelerator Science, Old Dominion University, Norfolk, Virginia 23529, USA.}

\author{Takayuki Kubo}
\email[]{kubotaka@post.kek.jp}
\affiliation{Department of Physics and Center for Accelerator Science, Old Dominion University, Norfolk, Virginia 23529, USA.}
\affiliation{KEK (High Energy Accelerator Research Organization), Tsukuba, Ibaraki 305-0801, Japan.}
\affiliation{SOKENDAI (the Graduate University for Advanced Studies), Hayama, Kanagawa 240-0115, Japan.}

\date{\today}

\begin{abstract}

We calculate a low-frequency surface impedance of a dirty, s-wave superconductor with an imperfect surface incorporating either a thin layer with a reduced pairing constant or a thin, proximity-coupled normal layer. Such structures model realistic surfaces of superconducting materials which can contain oxide layers, absorbed impurities or nonstoichiometric composition.  We solved the Usadel equations self-consistently and obtained  spatial distributions of the order parameter and the quasiparticle density of states which then were used to calculate a low-frequency surface resistance $R_s(T)$ and the magnetic penetration depth $\lambda(T)$ as functions of temperature in the limit of local London electrodynamics. It is shown that the imperfect surface in a single-band s-wave superconductor results in a non-exponential temperature dependence of $Z(T)$ at $T\ll T_c$ which can mimic the behavior of multiband or d-wave superconductors. The imperfect surface and the broadening of the gap peaks in the quasiparticle density of states $N(\epsilon)$ in the bulk give rise to a weakly temperature-dependent residual surface resistance.  We show that the surface resistance can be optimized and even reduced below its value for an ideal surface by engineering $N(\epsilon)$ at the surface using pairbreaking mechanisms, particularly, by incorporating a small density of magnetic impurities or by tuning the thickness and conductivity of the normal layer and its contact resistance. The results of this work address the limit of $R_s$ in superconductors at $T\ll T_c$, and the ways of engineering the optimal density of states by surface nano-structuring and impurities to reduce losses in superconducting micro-resonators, thin film strip lines, and radio frequency cavities for particle accelerators. 
\end{abstract}

\pacs{74.25. Nf, 74.45.+c, 74.78.Fk }

\maketitle

\section{Introduction}\label{section_introduction}

The physics of electromagnetic response of superconductors has been an area of active fundamental research relevant to many applications. 
For instance, quasi-particles generated due to absorption of infrared photons with energies higher than the gap energy $\hbar\omega > 2\Delta$ are essential for microwave kinetic inductance detectors 
of cosmic photons ~\cite{Zmuidzinas}. At low frequencies $\hbar \omega \ll 2\Delta$, single photons cannot break Cooper pairs, so the low-field surface impedance  
$Z=i\omega\mu_0\lambda+R_s$ is determined by the quasi-static London penetration depth $\lambda$ and the surface resistance $R_s$.
At temperatures $T$ well below the critical temperature $T_c$ and $\hbar\omega\ll \Delta$, s-wave superconductors have very small $R_s$ giving rise to extremely high quality factors instrumental for micro-resonators for quantum computing \cite{QC} or radio-frequency superconducting (SRF) cavities for particle accelerators \cite{Hasan_book}. The surface resistance in the Meissner state has the following 
generic temperature dependence observed in many experiments \cite{hein}:
\begin{equation}
R_s = (A\omega^2/T)\exp(-\Delta/k_BT) + R_i, \qquad \hbar\omega\ll\Delta.
\label{bcs}
\end{equation}
The first term in the right hand side of Eq. (\ref{bcs}) is the BCS surface resistance resulting from the RF heating of thermally-activated quasiparticles,  $A$ depends on purity of the material, and $\Delta = 1.76 k_BT_c$ is the superconducting energy gap \cite{mb,agh,nam,kopnin}.  The last term $R_i$ in Eq. (\ref{bcs}) is a residual surface resistance which sets the low-temperature limit of $R_s$.  For instance, typical values of $R_s\simeq 3-10~{\rm n\Omega}$ of the Nb resonator cavities operating  at $T=1.5\,{\rm K}$ and frequencies $\sim 1\,{\rm GHz}$ much smaller than the gap frequency $2\Delta/h \simeq 700\,{\rm GHz}$ exceed $R_s^{BCS}(T)$~ \cite{clare,agrev,padams}.  The residual resistance can also significantly exceed $R_s^{BCS}(T)$ in Nb$_3$Sn \cite{nb3sn1,nb3sn2,nb3sn3,nb3sn4}, MgB$_2$ \cite{mgb2a,mgb2b,mgb2c}, and iron-based superconductors \cite{fbs1,fbs2,fbs3,fbs4}.  

Mechanisms of the residual surface resistance are not well understood, but it has been established that $R_i$ can be changed significantly by the materials treatment or by irradiation \cite{halama}. For instance, lossy oxides or metallic hydrides at the surface of Nb,  grain boundaries or trapped vortices which appear during cooldown of the sample through $T_c$, surface roughness and segregation of impurities at the surface can contribute to $R_i$~ \cite{clare,agrev,padams}. These extrinsic factors can be ameliorated by the materials treatments, and by pushing out a fraction of trapped vortices by strong temperature gradients \cite{flux1,flux2,flux3}, so the fundamental lower limit of $R_i$ is of great interest.

A finite $R_i$ in the Meissner state does not come from the BCS model in which the quasiparticle density of states (DOS) $N(\epsilon)$ vanishes at all energies $|\epsilon|<\Delta$ even in the presence of weak nonmagnetic impurities \cite{balatski}.  However, numerous tunneling experiments have shown that in the observed $N(\epsilon)$ the BCS singularities at $\epsilon=\Delta$ are smeared out and subgap states with finite $N(\epsilon)$ appear at $|\epsilon|<\Delta$. Such $N(\epsilon)$ has been often described by the phenomenological Dynes formula \cite{dynes1,dynes2}:
\begin{equation}
N(\epsilon) = \mbox{Re}\frac{N_s(\epsilon+i\Gamma )}{\sqrt{(\epsilon+i\Gamma )^2 -  \Delta^2}}.
\label{NE}
\end{equation}
Here the damping parameter $\Gamma$ quantifies a finite lifetime of quasiparticles $\sim \hbar/\Gamma$, and $N_s$ is the density of states at $T>T_c$.  Detailed discussions of tunneling measurements of $N(\epsilon)$ and application of Eq. (\ref{NE}) can be found in Ref. \onlinecite{tun}. Different mechanisms of broadening of DOS peaks have been considered in the literature, including inelastic scattering of quasiparticles on phonons \cite{kopnin,inelast}, strong Coulomb correlations \cite{coulomb}, anisotropy of the Fermi surface \cite{anis}, local inhomogeneities of the BCS pairing constant \cite{larkin}, magnetic impurities \cite{balatski,glaz,khariton}, and effects of spatial correlations in impurity scattering \cite{balatski,meyer}. 

The broadening of DOS peaks can result in a non-exponential dependence of $R_s(T)$ and the leveling off the Arrhenius plot of $\ln R_s$ versus $1/T$ at low temperatures \cite{ag_sust}, which is usually attributed to a residual resistance. Indeed, Eq. (\ref{NE}) suggests a finite density of states $N_s\Gamma/\Delta$ at the Fermi level, which would cause a finite $R_i$ at $T=0$.  This was shown for the dirty limit \cite{ag_sust} and then extended to an arbitrary impurity concentration \cite{hlub}.  Yet  Eq. (\ref{NE}) has not been derived from the microscopic theory of superconductivity, so not only the dependencies of $\Gamma$ on $T$ and $\epsilon$ but also the validity of Eq. (\ref{NE}) at $T\ll T_c$ remain unclear. For instance, spatial correlations in impurity scattering can result in an exponential low-energy tail in $N(\epsilon)$ \cite{balatski}, and any power-law temperature dependence of $\Gamma(T)$ would manifest itself as an apparent residual resistance in the Arrhenius plot measured within a limited temperature window. 

The link between the subgap states and the residual resistance \cite{ag_sust} suggests that both $R_i$ and $N(\epsilon)$ can be very sensitive to the state of the surface. Indeed, tunneling measurements of $N(\epsilon)$ are often masked by metallic suboxies, local reduction of the BCS pairing constant, absorbed impurity layers or surface nonstoichiometry which can weaken superconductivity at the surface \cite{tun}. The importance of the surface contributions to the tunneling DOS was recognized long ago \cite{appelbaum,mcmillan}, but the extent to which $\Gamma$ in Eq. (\ref{NE}) represents a true bulk value or it is mostly controlled by the surface properties is not well understood. Yet the exponentially small surface resistance at $T\ll T_c$ becomes extremely sensitive to any surface contributions which yield a non-exponential temperature dependence of $R_s(T)$. The fact that the observed values of $R_i$ in Nb could be accounted for by rather small $\Gamma\simeq (0.02-0.05)\Delta$~\cite{ag_sust} suggests that, if $\Gamma$ does come from the surface effects, superconductivity is weakened in a surface layer thinner than the coherence length $\xi$. This conclusion is consistent with the well-established structure of the Nb surface covered by a layer of dielectric ${\rm Nb_2 O_5}$ oxide followed by the layer of normal (N) metallic sub-oxide ${\rm NbO}$ and a dirty Nb superconducting (S) layer in which the order parameter is reduced by diffused oxygen impurities \cite{clare,agrev,padams}. The thickness of the suboxide layer $\simeq 1-2$ nm is much smaller than $\xi\simeq 40$ nm, so this layer becomes superconducting due to the proximity effect. This structure is characteristic of the  surface of many superconducting materials, particularly Nb$_3$Sn, MgB$_2$ or iron-based superconductors, which can also exhibit a significant surface nonstoichiometry.   

In this work we calculate $Z(\omega,T)$ for a realistic surface modeled by a thin layer of weakened superconductivity or by N layer coupled to the bulk supercondnuctor by the proximity effect. This model allows us to calculate $Z(\omega,T)$ using the  well-developed approach based on the Usadel equations  \cite{Belzig_review,golubov_rev} for the proximity-coupled dirty N-S bilayers.  Previous calculations of such N-S bilayers have shown significant broadening of DOS peaks and low-energy minigaps in the N layer~ \cite{proxi1,proxi2} which can manifest itself in dc screening and non-exponential temperature dependencies of $R_s(T)$ and $\lambda(T)$ at $T\ll T_c$. Screening of a dc parallel field and magnetic breakdown of superconductivity in N-S bilayers has been thoroughly investigated theoretically for arbitrary mean free path and temperatures \cite{dc1,dc2,dc3,dc4}. In turn, the nonexponential temperature dependence of $R_s(T)$ was observed in microwave measurements on S-N bilayers of different materials~ \cite{anlage1,anlage2,anlage3,anlage4}. However, unlike the non-dissipative dc magnetic response, the surface resistance is rather sensitive to the details of peaks and low-energy tails in $N(\epsilon)$, so a theory of $R_s(T,\omega)$ should include self-consistent calculations 
of spatial variation of DOS perpendicular to the surface. In this work we develop such a theory of $R_s$ which incorporates both bulk and surface subgap states, and the residual surface resistance into the BCS theory of electromagnetic response \cite{nam}. Since a moderate broadening of DOS peaks can reduce the low-frequency $R_s(T)$ at intermediate temperatures \cite{ag_sust}, this theory also shows how $R_s$ could be optimized using pairbreaking mechanisms, for example by tuning the concentration of paramagnetic impurities or properties of N layer at the surface. Such engineering of an optimum DOS by surface nanostructuring and impurity management may be used for the material optimization and increasing the quality factors of the SRF resonant cavities for particle accelerators and microresonators for quantum information processing and photon detectors.

The paper is organized as follows. In Sec. II we formulate the Usadel equations for the quasi-classical Green's functions and the boundary conditions. In Sec. III we solve the Usadel equations for a semi-infinite superconductor covered by a thin layer with a reduced pairing constant $g(x)$, and a thin proximity-coupled N layer at the surface. For both cases, we calculate self-consistently the Green's functions, $\Delta(x)$, $N(\epsilon,x)$ and show that the effect of a thin surface layer extends into the bulk over distances much larger than $\xi$ for quasiparticle energies close to $\Delta$. In Sec. IV we calculate the effect of the surface layer on the magnetic penetration depth and the surface resistance in the local London limit. It is shown that the non-ideal surface can result in a non-exponential dependencies of $\lambda(T)$ and $R_s(T)$ at low temperatures which can extend down to very low $T\ll T_c$ for a large S-N interface resistance, and $R_s$ can be minimized by tuning the properties of N layer. In Sec. V we discuss implications of the obtained results for the interpretation of experimental data on the measurements of surface impedance at low temperatures, and optimization of $R_s$.

\section{Usadel equations and boundary conditions}\label{section_theory}

%
\begin{figure}[tb]
   \begin{center}
   \includegraphics[scale=0.35,trim={60mm 0 0 0},clip]{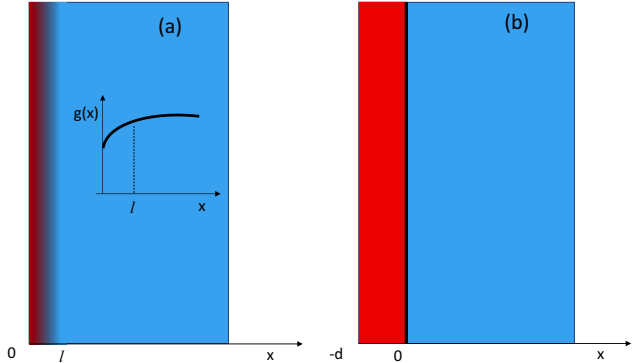}
   \end{center}\vspace{-12mm}
   \caption{ (a) A surface layer of gradually reduced BCS pairing constant $g(x)$. Inset shows a profile of $g(x)$. (b) A superconductor covered with a normal layer of thickness $d$. 
   The vertical black line in (b) shows the S-N interface giving rise to the contact resistance $R_B$. 
   }\label{fig1}
\end{figure}

Consider the geometry shown in Fig.~\ref{fig1} which represents a superconductor with a thin surface layer ($0\le x \lesssim \ell$) of reduced BCS pairing constant $g$, and a superconductor ($x\ge 0$) covered with a thin N layer ($-d \le x <0$).  We use the quasicassical theory ~\cite{Belzig_review,golubov_rev} for a dirty superconductor described by the normal and anomalous thermodynamic Green functions $G = \cos\theta$ and $F = \sin\theta$, where $\theta(x)$ obeys the Usadel equation:  
\begin{equation}
\frac{\hbar D}{2}\theta'' = -\Delta(x) \cos\theta + \hbar \omega_n \sin\theta.
\label{eq:im_Usadel_1}
\end{equation}
Here $D$ is the electron diffusivity, and the prime denotes differentiation with respect to $x$.  
The pair potential $\Delta(x)$ satisfies the self-consistency equation
\begin{equation}
\Delta(x) = 2\pi k_B T g(x) \sum_{\omega_n>0}^{\Omega} \sin\theta(x), 
\label{eq:self-consistency}
\end{equation}
where the summation over the Matsubara frequencies
$ \omega_n = \pi k_B T (2n+1)/\hbar$ is cut off at the Debye frequency $\Omega$. Equation (\ref{eq:im_Usadel_1}) written for both N and S regions are supplemented by the boundary conditions at the outer surface 
and $x\to\infty$:
\begin{eqnarray}
\theta'|_{\rm surface}=0 ,  \label{eq:BC_-d} \\
\theta (\infty) = \theta_{\infty} , \label{eq:BC_Inf}
\end{eqnarray}
Here $\theta_{\infty}$ defines the uniform Green functions:
\begin{eqnarray}
G(\omega_n)=\cos\theta_{\infty}= \frac{\hbar \omega_n}{\sqrt{(\hbar\omega_n)^2+\Delta^2}},\\
\label{cos}
F(\omega_n)=\sin\theta_{\infty}= \frac{\Delta}{\sqrt{(\hbar\omega_n)^2+\Delta^2}}, 
\label{sin}
\end{eqnarray}
where $\Delta\equiv \Delta(\infty)$ denotes an equilibrium order parameter in S region far away from the surface, as opposed to the 
varying pair potential $\Delta(x)$ in Eq. (\ref{eq:self-consistency}).
For the case shown in Fig.~\ref{fig1}(b), we use the following boundary conditions at the S-N interface ~\cite{Kuprianov_Lukichev}:
\begin{eqnarray} 
&&\sigma_nR_B\theta_{-}' = \sin (\theta_{0} - \theta_{-}) , 
\label{eq:im_BC_gammaB} \\
&&\sigma_n \theta_{-}' = \sigma_s \theta_{0}' ,
\label{eq:im_BC_gamma} 
\end{eqnarray}
Here $\theta_{-} = \theta|_{x=-0}$, $\theta_0 = \theta|_{x=+0}$,  $R_B$ is the N-S contact resistance, and $\sigma_n$ and $\sigma_s$ are the normal state conductivities in N and S regions, respectively. 
It is convenient to define the following dimensionless parameters:
\begin{eqnarray}
&&\alpha =\frac{N_n}{N_s} \frac{d}{\xi_S} , 
\label{alpha} \\
&&\beta= \frac{4e^2}{\hbar}R_B N_n \Delta d, 
\label{beta}
\end{eqnarray}
where $N_n$ and $N_s$ are the normal densities of states in N and S regions, and $\xi_S$ and $\xi_N$ are the 
respective coherence lengths in the dirty limit:
\begin{equation}
\xi_N=\sqrt{\frac{\hbar D_n}{2\Delta}}, \qquad \xi_S=\sqrt{\frac{\hbar D_s}{2\Delta}}.
\label{xi}
\end{equation}
Notice that $\alpha$ and $\beta$ are independent of the mean free path in the N layer.  If $N_n=N_s$, we have $\alpha=d/\xi_S=0.05$ for a moderately dirty Nb with 
$\xi_S=20\,{\rm nm}$ covered by N layer of thickness $d=1\,{\rm nm}$. 

Equations (\ref{eq:im_BC_gammaB})-(\ref{eq:im_BC_gamma}) result from the general boundary conditions \cite{nazarov,eschrig} for quasiclassical 
Green's functions if the N-S interface has a small transmission coefficient $t\sim \pi R_K/R_Bk_F^2 \ll 1$, where $k_F$ is the Fermi wave vector, and $R_K=h/e^2$.  The condition $t\ll 1$ imposes a restriction 
on $\beta$ which becomes apparent by presenting Eq. (\ref{beta}) in the form, $\beta=(4d/\pi\xi_0)(k_F^2R_B/\pi R_K)$, where $N_n=m^*k_F/2\pi^2\hbar^2$, $m^*$ is the electron effective mass, and $\xi_0=\hbar v_F/\pi\Delta$ is a coherence length in the clean limit. Thus, the parameter $\beta$ at $t\ll 1$ is confined within the region $d/\xi_0\ll \beta < \infty$, which at $d/\xi_0\ll 1$ comprises the essential cases of both $\beta>1$ and $\beta\ll1$ considered in this work. As a result,  Eqs. (\ref{eq:im_BC_gammaB})-(\ref{eq:im_BC_gamma}) can be used for the calculations of $R_s$ for a thin,   
proximity-coupled N layers with both $\beta\ll 1$ and $\beta>1$, and qualitatively for intermediate transparency $t\sim 1$ and $\beta \ll d/\xi_0$.       

Retarded Green functions $G^R = \cosh\theta$ and $F^R = \sinh\theta$ are obtained by solving the Usadel equation in the real-frequency representation $\hbar\omega \to -i(\epsilon+i\Gamma)$:
\begin{equation}
\frac{\hbar D}{2}\theta'' = i \Delta(x)\cosh\theta - i (\epsilon+i\Gamma) \sinh\theta, 
\label{eq:Usadel_1}
\end{equation}
where $\Delta(x)$ satisfies Eq.~(\ref{eq:self-consistency}), and $\Gamma$ accounts for a finite quasiparticle lifetime. 
For a uniform superconductor, 
\begin{eqnarray}
&&G^R(\epsilon)=\cosh\theta_{\infty}=
\frac{\epsilon + i\Gamma}{\sqrt{(\epsilon + i\Gamma)^2-\Delta^2}} , \label{eq:cosh_inf_Dynes} \\
&&F^R(\epsilon)=\sinh\theta_{\infty}=
\frac{\Delta}{\sqrt{(\epsilon + i\Gamma)^2-\Delta^2}} .
\end{eqnarray}
The self-consistency equation for $\Delta$  in the bulk is obtained by substituting Eq. (\ref{sin}) into Eq. (\ref{eq:self-consistency}):
\begin{equation}
\!\!\ln\frac{T}{T_{c0}}=2\pi k_B T\sum_{\omega_n} \left[\frac{1}{\sqrt{(\hbar\omega_n+\Gamma)^2+\Delta^2}}-\frac{1}{\hbar\omega_n}\right]\!,
\label{gapg}
\end{equation}
where $T_{c0}=(2\tilde{\gamma}\hbar\Omega/\pi k_B)\exp(-1/g)$ is the BCS critical temperature at $\Gamma=0$, and $\ln\tilde{\gamma}=\gamma_E=0.577$ is the Euler constant. Setting $\Delta=0$ and summing over $\omega_n$ 
in Eq. (\ref{gapg}) yields the following equation for $T_c$:
\begin{equation}
\ln\frac{T_c}{T_{c0}}+\psi\left(\frac{1}{2}+\frac{\Gamma}{2\pi k_BT_c}\right)-\psi\left(\frac{1}{2}\right)=0,
\label{tcg}
\end{equation}
where $\psi(z)$ is a digamma function. Equation (\ref{tcg}) has the form of the well-known equation for $T_c(\Gamma)$ of a superconductor with paramagnetic impurities \cite{balatski,maki}. Here $T_c$ decreases with $\Gamma$ and vanishes at $\Gamma=\Delta_0/2$, where $\Delta_0=2\hbar\Omega\exp(-1/g)$ is the BCS gap at $T=0$ and $\Gamma=0$. 
The bulk pair potential $\Delta(\Gamma)$ at low temperatures $T\ll T_c$ can be obtained by replacing the summation in Eq. (\ref{gapg}) with integration over $\omega$, which yields  
$\Delta^2=\Delta_0(\Delta_0-2\Gamma)$. In the case of $\Gamma\ll\Delta_0$ considered in this work,  $T_c$ and $\Delta$ take the form:  
\begin{gather}
T_c=T_{c0}-\frac{\pi\Gamma}{4k_B}, 
\label{tcg1} \\
\Delta = \Delta_0-\Gamma.
\label{delg} 
\end{gather}
In the  following we assume no BCS pairing in the N layer, adopt $\xi_S$ and $\Delta$ as units of length and energy and use the dimensionless parameters  
$x/\xi_S$, $k_B T/\Delta$, $\hbar\omega/\Delta$, and $\epsilon/\Delta$ unless stated otherwise.

\section{Solution of the Usadel equation} 

\subsection{Surface layer with a reduced pairing constant} \label{section:simple_ex}

\subsubsection{Self-consistent pair potential} 

Consider a thin surface layer with inhomogeneous BCS pairing constant $g(x)=g+\delta g(x)$, as shown in Fig.~\ref{fig1}(a), where $g=g(\infty)$. For a weak perturbation $\delta g(x)\ll g$,
resulting in a weak disturbance of $\theta(x) = \theta_{\infty} + \delta\theta(x)$ and $\Delta(x) = 1 + \delta \Delta(x)$, 
the linearized thermodynamic Usadel equation for $\delta\theta\ll\theta_\infty$ takes the form
\begin{equation}
\delta\theta'' - k_{\omega}^2 \delta\theta = - \cos\theta_{\infty} \delta\Delta(x), 
\label{eq:ex_thermodynamic_Usadel}
\end{equation}
where $k_{\omega} = (\omega_n^2 +1)^{\frac{1}{4}}$ and $\delta\Delta(x)$ satisfies the linearized gap equation  
\begin{equation}
\delta \Delta(x) = 2\pi T \sum_{\omega}^{\Omega} [g\cos\theta_{\infty} \delta\theta(x) + \delta g(x) \sin\theta_{\infty}].
\label{eq:ex_self_consistency}
\end{equation}
Equations (\ref{eq:ex_thermodynamic_Usadel}) and (\ref{eq:ex_self_consistency}) can be solved by 
the cosine Fourier transform as shown  Appendix~\ref{appendix:simple_ex}:
\begin{eqnarray}
&&\delta\Delta(x) = \frac{1}{\pi g^2}\int_0^{\infty} \!\!dk \frac{\delta g_k}{S(k)} \cos kx ,
\label{eq:ex_deltaDelta_sol} 
\\
&&S(k) = 2\pi T \sum_{\omega}^{\infty} \frac{k^2\sqrt{\omega_n^2+1}+1}{(\omega_n^2+1)(k^2+\sqrt{\omega_n^2+1})},
\end{eqnarray}
where $\delta g_k$ is the Fourier image of $\delta g(x)$. Notice that $S(k)$ is a slow function of $k$, varying from $S(k)=1+\pi k^2/4$ at $k^2\ll 1$ to $S(k)=\ln( 2 k^2)$ at $k^2 \gg 1$. 
As an illustration, consider $\delta \Delta(x)$ for the exponential profile of $\delta g(x) = -\zeta g \exp(-x/\ell)$ for which:
\begin{equation}
\frac{\delta g_k}{g} = -\frac{ \ell \zeta}{1+k^2 \ell^2}, 
\label{eq:delta_g_k}
\end{equation}
where the parameter $\zeta < 1$ quantifies the magnitude of $\delta g(x)$. 
Since $S(k)\sim 1$ varies very slowly with $k$ at $k\gg 1$, the integral in  Eq.~(\ref{eq:ex_deltaDelta_sol}) converges at $k\sim 1/\ell$. 
Thus, the disturbance $\delta \Delta(x) \propto \delta g(x)$ decays over a short length $\sim\ell \ll \xi_S$ 
much smaller than the length scales of variation of the retarded Green functions, as will be shown below.

\subsubsection{Retarded Green functions and density of states} \label{reducedBCS_retarded}

To calculate the retarded Green functions we solve the Usadel equation in the real-frequency representation, 
\begin{equation}
\theta''(x) = i[1+\delta\Delta(x)]\cosh\theta(x) -  i \epsilon \sinh\theta(x) .  \label{eq:reducedBCS_real_Usadel}
\end{equation}
Because the disturbance of the pair potential $\delta \Delta(x) \propto \delta g(x)$ decreases rapidly over the length $\ell \ll \xi_S$, we can approximate  
$\delta \Delta (x)$ as follows:
\begin{equation}
\delta \Delta(x) = -\Psi \delta(x),
\label{deldel}  
\end{equation}
where $\Psi$ is given by the Fourier component $\delta\Delta_k$ at $k=0$: 
\begin{equation}
\Psi = \frac{1}{g^2}\int_0^\infty\delta g(x)dx = \frac{\zeta \ell}{g}.
\end{equation}
The solution of Eq.~(\ref{eq:reducedBCS_real_Usadel}) is given by (see Appendix~\ref{appendix:reducedBCS_retarded})  
\begin{equation}
\tanh\frac{\theta(x)-\theta_{\infty}}{4} = \tanh \frac{\theta_0-\theta_{\infty}}{4}  e^{-k_{\epsilon}x} , \label{eq:reducedBCS_solution}
\end{equation}
where $k_{\epsilon}\equiv (1-\epsilon^2)^{1/4}$. The value $\theta_0\equiv \theta(x=0)$ is determined by a self-consistency equation which is obtained by multiplying 
Eq.~(\ref{eq:reducedBCS_real_Usadel}) by $\theta'$ and integrating from $x=+0$ to $\infty$ using the boundary conditions $\theta'(+0)=-i\Psi\cosh\theta_0$, $\theta(\infty)=\theta_\infty$, and  
$\theta'(\infty)=0$. As shown in Appendix B, this procedure yields the following equation for $\theta_0$: 
\begin{equation}
2 k_{\epsilon} \sinh \frac{\theta_0 -\theta_{\infty}}{2} =i\Psi \cosh\theta_0.
\label{eq:reducedBCS_theta0}
\end{equation}
The solutions for the Green functions are then 
\begin{gather}
\!\!\!\!\!G^R= \frac{4t(1+t^2)}{(1-t^2)^2}\sinh\theta_{\infty} +\! \biggl[ \frac{2(1+t^2)^2}{(1-t^2)^2} -1 \biggr]\!\cosh\theta_\infty , \label{GR} \\
\!\!\!\!\!F^R = \frac{4t(1+t^2)}{(1-t^2)^2}\cosh\theta_{\infty} +\!\biggl[ \frac{2(1+t^2)^2}{(1-t^2)^2} -1 \biggr]\!\sinh\theta_\infty ,  \label{FR}
\end{gather}
where $t(x)=\tanh[(\theta(x)-\theta_\infty)/4]$ is given by:
\begin{equation}
t(x) =\biggl( \tanh \frac{\theta_0-\theta_{\infty}}{4}  \biggr) e^{-k_{\epsilon} x}  .
\label{t}
\end{equation} 
Equations (\ref{eq:reducedBCS_theta0})-(\ref{FR}) define self-consistently $\theta(x)$ in a 
superconductor with a thin pair breaking layer. 

\begin{figure}[tb]
   \begin{center}
   \includegraphics[width=1\linewidth]{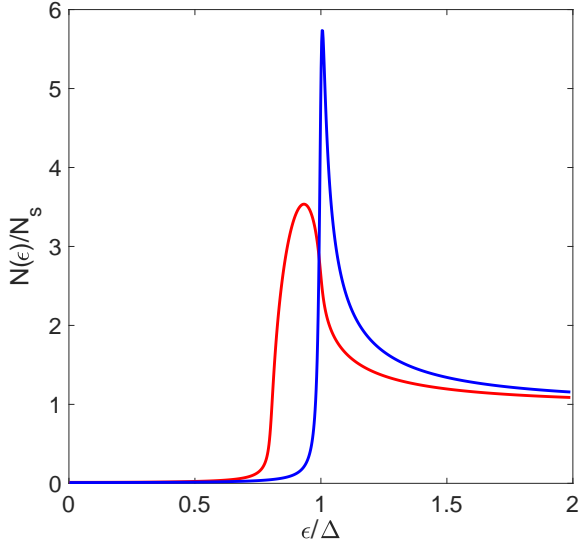}
   \end{center}\vspace{0cm}
   \caption{
Density of states at the surface calculated for $\Psi=0.2$ and $\Gamma=0.01$ (red line). The blue line shows DOS in the bulk. 
   }\label{fig2}
\end{figure}
Equations (\ref{eq:reducedBCS_solution})-(\ref{t}) allow us to calculate the effect of the pairbreaking layer on the normalized DOS,  $n(\epsilon)= N(\epsilon)/N_s ={\rm Re}[G^{R}(\epsilon,x)]$.
Shown in Fig.~\ref{fig2} is  $n(\epsilon, x=0) = {\rm Re}\cosh\theta_0$ at the surface  calculated from Eq. (\ref{eq:reducedBCS_theta0}) at $\Psi=0.2$ and $\Gamma/\Delta=0.01$, along with the Dynes DOS with $\Gamma/\Delta=0.01$ in the bulk. The surface pair breaking layer broadens the peak in DOS and shifts it to lower energies, which may complicate extraction of the bulk gap $\Delta$ from tunneling measurements using the conventional fitting procedure based on Eq. (\ref{NE}).

\subsection{Effect of normal layer at the surface.} 

\subsubsection{Self-consistent pair potential}\label{NS_Matsubara}

Consider N layer of thickness $d <\xi_N$ at the surface, as shown in Fig.~\ref{fig1}(b). 
To calculate $\theta(x)$ and $\Delta(x)$, 
we first solve the thermodynamic Usadel equation  
$\xi_N^2 \theta'' = \omega_n \sin\theta$ in the N layer at $-d \le x < 0$. 
Since $\theta(x)$ varies weakly over the thin N layer with $d \ll \xi_N$, the solution 
satisfying the boundary condition $\theta'(-d)=0$ can be approximated by %
\begin{equation}
\theta(x) = \theta_{-} + \frac{x(x+2d)}{2\xi_N^2}\omega_n  \sin\theta_{-}. 
\label{eq:im_theta_M}
\end{equation}
The relation between the boundary values $\theta_-$ and $\theta_0$ at the N and S sides of the interface can be obtained using Eq.~(\ref{eq:im_BC_gammaB}) and $\theta'_{-}=(d/\xi_N^2)\omega_n \sin\theta_{-}$: 
\begin{equation}
\sin \theta_{-} = \frac{\sin\theta_0}{\sqrt{1+\beta^2\omega_n^2 + 2\beta \omega_n\cos\theta_{0}}}.
\label{sin-}
\end{equation}
Then Eq.~(\ref{eq:im_BC_gamma}) becomes
\begin{gather}
\theta'(0) = \omega_n \Phi \sin\theta_{0} , \label{eq:BC_Phi}  \\
\Phi =  \frac{\alpha}{\sqrt{1+\beta^2\omega_n^2 + 2\beta \omega_n\cos\theta_{0}}} . \label{eq:Phi_def} 
\end{gather}
Now the problem is reduced to solving the Usadel equation in S region:
\begin{equation}
\theta'' = -[1+\delta\Delta(x)]\cos\theta +  \omega_n\sin\theta,
\label{ust}
\end{equation} 
with the boundary condition (\ref{eq:BC_Phi}) at $x=0$, where  
$\delta\Delta(x)$ is a short-range perturbation of the pair potential  approximated by Eq. (\ref{deldel}) with the amplitude $\Psi$ to be calculated self-consistently. 
The solution satisfying the boundary condition (\ref{eq:BC_Inf}) is given by (see Appendix~\ref{appendix:NS_im_Usadel})
\begin{equation}
\tan\biggl[\frac{\theta(x)-\theta_{\infty}}{4}\biggr] = \tan \bigg[\frac{\theta_0-\theta_{\infty}}{4}\biggr] e^{-k_{\omega}x} . \label{eq:NS_solution}
\end{equation}
Here $\theta_0$ and $\Psi$ satisfy the following equations:
\begin{eqnarray}
2 k_{\omega} \sin \frac{\theta_0 -\theta_{\infty}}{2} + \omega_n \Phi \sin\theta_0  + \Psi \cos \theta_0 =0, \label{eq:NS_theta0}
\\
\Psi = 2\pi T g \sum_{\omega_n>0}^\Omega \int_0^{\infty}[\sin\theta_\infty-\sin\theta(x)]dx. 
\label{eq:NS_self-consistency}
\end{eqnarray} 

\begin{figure}[tb]
   \begin{center}
   \includegraphics[width=1\linewidth]{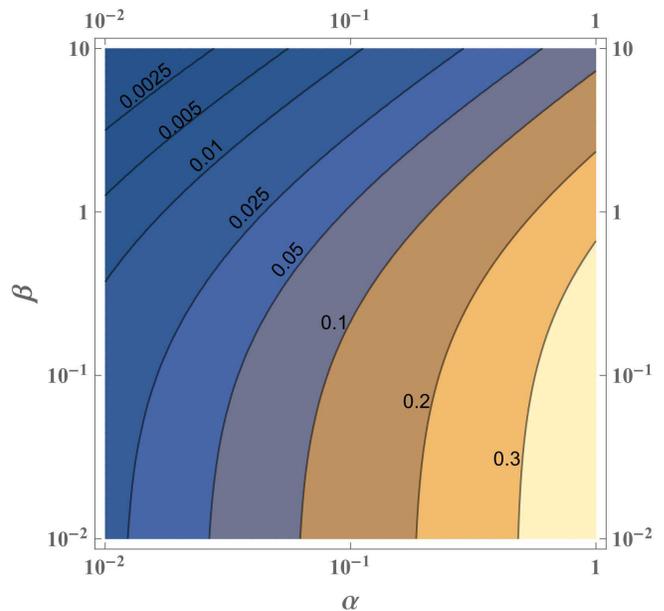}
   \end{center}\vspace{0cm}
   \caption{
A contour map of the self-consistent pair potential $\Psi$  calculated from Eq.~(\ref{eq:self_consistency_gamma2}) at $k_BT/\Delta=0.057$ and $\hbar\Omega = 11\Delta$. 
The parameters $\alpha$  and $\beta$ are defined by Eqs. (\ref{alpha}) and (\ref{beta}). 
   }\label{fig3}
\end{figure}

The closed set of Eqs. (\ref{eq:Phi_def}) and (\ref{eq:NS_solution})-(\ref{eq:NS_self-consistency}) simplifies in the limit 
of $\Psi \ll 1$ which encompasses a range of the parameters of practical interest. 
In this case Eqs.~(\ref{eq:NS_solution}) and (\ref{eq:NS_theta0})  can be  linearized in $\delta\theta = \theta-\theta_{\infty} \ll 1$, giving
\begin{equation}
\delta\theta(x) = \delta\theta_0 e^{-k_{\omega} x} , \qquad\quad
\delta\theta_0 = -\frac{\omega_n (\Phi+\Psi)}{k_{\omega}^3+\Phi\omega_n^2} .
\label{eq:Matsubara_deltaTheta} 
\end{equation}
Substituting Eq.~(\ref{eq:Matsubara_deltaTheta}) into Eq.~(\ref{eq:NS_self-consistency}) yieds:
\begin{equation}
\Psi = \frac{S_{\rm N}}{1-S_{\rm D}} ,
\label{eq:self_consistency_gamma2} \\
\end{equation}
where 
\begin{gather}
S_{\rm N} = 2\pi T g \sum_{\omega_n>0}^{\Omega} \frac{\omega_n^2 \Phi}{k_{\omega}^3 (k_{\omega}^3+\Phi \omega_n^2) } , 
\label{eq:SN} \\
S_{\rm D} = 2\pi T g \sum_{\omega_n>0}^{\Omega} \frac{\omega_n^2}{k_{\omega}^3 (k_{\omega}^3+\Phi \omega_n^2) } .\label{eq:SD}
\end{gather}
Figure~\ref{fig3} shows a contour plot of $\Psi$ calculated from Eqs.~(\ref{eq:Phi_def}) and (\ref{eq:Matsubara_deltaTheta})-(\ref{eq:SD}), from which it follows that     
the condition $\Psi < 1$ is satisfied in a wide range of $\alpha$ and $\beta$. 
At $T\ll T_c$, the summation in Eqs. (\ref{eq:SN}) and (\ref{eq:SD}) can be replaced by integration, 
allowing for analytical evaluation of $\Psi$ in certain cases summarized in Appendix~\ref{appendix:SN_SD}.
For instance, in the practically important case of $\alpha \ll 1$ and $\beta >\alpha^2/4$, we have:
\begin{gather}
\Psi=\frac{\alpha(\beta-1)}{1+\beta^2} +
\nonumber \\
\frac{\alpha}{(1+\beta^2)^{3/2}}\ln\frac{(1+\beta\Lambda)(\beta+\sqrt{1+\beta^2})}{\sqrt{(1+\Lambda^2)(1+\beta^2)}-\Lambda+\beta},
\label{Psi}
\end{gather}
where $\Lambda = \hbar\Omega/\Delta$ is a large parameter of the BCS model. Yet for real materials, $\Lambda$ may not be necessarily the largest parameter in Eq. (\ref{Psi}): for Nb with $\Delta\simeq 17.5$ K and $\hbar\Omega/k_B\simeq 184$ K, we obtain $\Lambda\simeq 10.5$, so the BCS limit of $\Lambda\gg 1$ should be taken with care. Indeed, for $\alpha=0.05$ used in our numerical simulations presented below, Eq. (\ref{Psi}) describes both cases of $\Lambda\beta\ll 1$ and $\beta\Lambda\gg 1$. If $\beta\Lambda\gg 1$, we have $\Psi\approx 0.62\alpha$ at $\beta=1$, $\Psi=\alpha[\ln(2/\beta)-1]$, at  $\beta\ll 1$, and $\Psi  = \alpha/\beta$ at $\beta\gg 1$. However if $\beta\Lambda\ll 1$,  Eq. (\ref{Psi}) shows that $\Psi$ becomes independent of $\beta$:
\begin{equation}
\Psi=\alpha\left(\ln\frac{2\hbar\Omega}{\Delta}-1\right)=\alpha\left(\frac{1}{g}-1\right).
\label{psss}
\end{equation}
Here the BCS gap equation $\Delta=2\hbar\Omega\exp(-1/g)$ at $T=0$ was used. For Nb, the conditions under which Eq. (\ref{psss}) is valid, overlaps with the condition $\alpha\ll 1$ under which the thin N-layer approximation of this work is applicable. Our numerical simulations for $\alpha=0.01$ and $\hbar\Omega=11\Delta$ show that Eq. (\ref{Psi}) describes the full Eq. (\ref{eq:self_consistency_gamma2}) 
to the accuracy better than $3\%$ in a wide range $10^{-2}\ <\beta<10^2$.

%

\subsubsection{Retarded Green functions and density of states}

Retarded Green's functions are obtained by solving the real-frequency Usadel equations in N and S regions. 
In N region $(x<0)$ we have: 
\begin{eqnarray}
&&G^R = \cosh \biggl[ \theta_{-} - \frac{i\epsilon x(x+2d)}{2\xi_N^2}\sinh\theta_{-} \biggr] , \label{eq:NS_GR_Nx}  \\
&&F^R = \sinh \biggl[ \theta_{-} - \frac{i\epsilon x(x+2d)}{2\xi_N^2} \sinh\theta_{-} \biggr] , \label{eq:NS_FR_Nx} 
\end{eqnarray}
where 
\begin{equation}
\sinh\theta_{-}  
= \frac{\sinh\theta_{0}}{\sqrt{1 - \beta^2\epsilon^2-2i\beta \epsilon \cosh\theta_{0}}} . \label{eq:NS_FR_N}
\end{equation}
Te boundary condition at $x=0$ is then, 
\begin{eqnarray}
&& \theta'(0) = -i\epsilon \Phi \sinh\theta_0 , \label{eq:NS_BC_interface} \\
&& \Phi \equiv \frac{\alpha}{\sqrt{1 - \beta^2\epsilon^2-2i\beta \epsilon \cosh\theta_{0}}} . \label{eq:NS_phi_real}
\end{eqnarray}
At $x\ge 0$, the Green functions are given by Eqs. (\ref{GR})-(\ref{t}) and 
$\theta_0$ satisfies the self-consistency equation:
\begin{equation}
2 k_{\epsilon} \sinh \frac{\theta_0 -\theta_{\infty}}{2} =i \epsilon \Phi \sinh\theta_0  +i \Psi \cosh \theta_0  \label{eq:NS_real_theta0}
\end{equation}
which takes into account the proximity effect in N layer and a reduction of the pair potential in S region, where  
$\Psi$ is given by Eq.~(\ref{eq:self_consistency_gamma2}) (see also Fig.~\ref{fig3}). 

\begin{figure}[tb]
   \includegraphics[width=9cm]{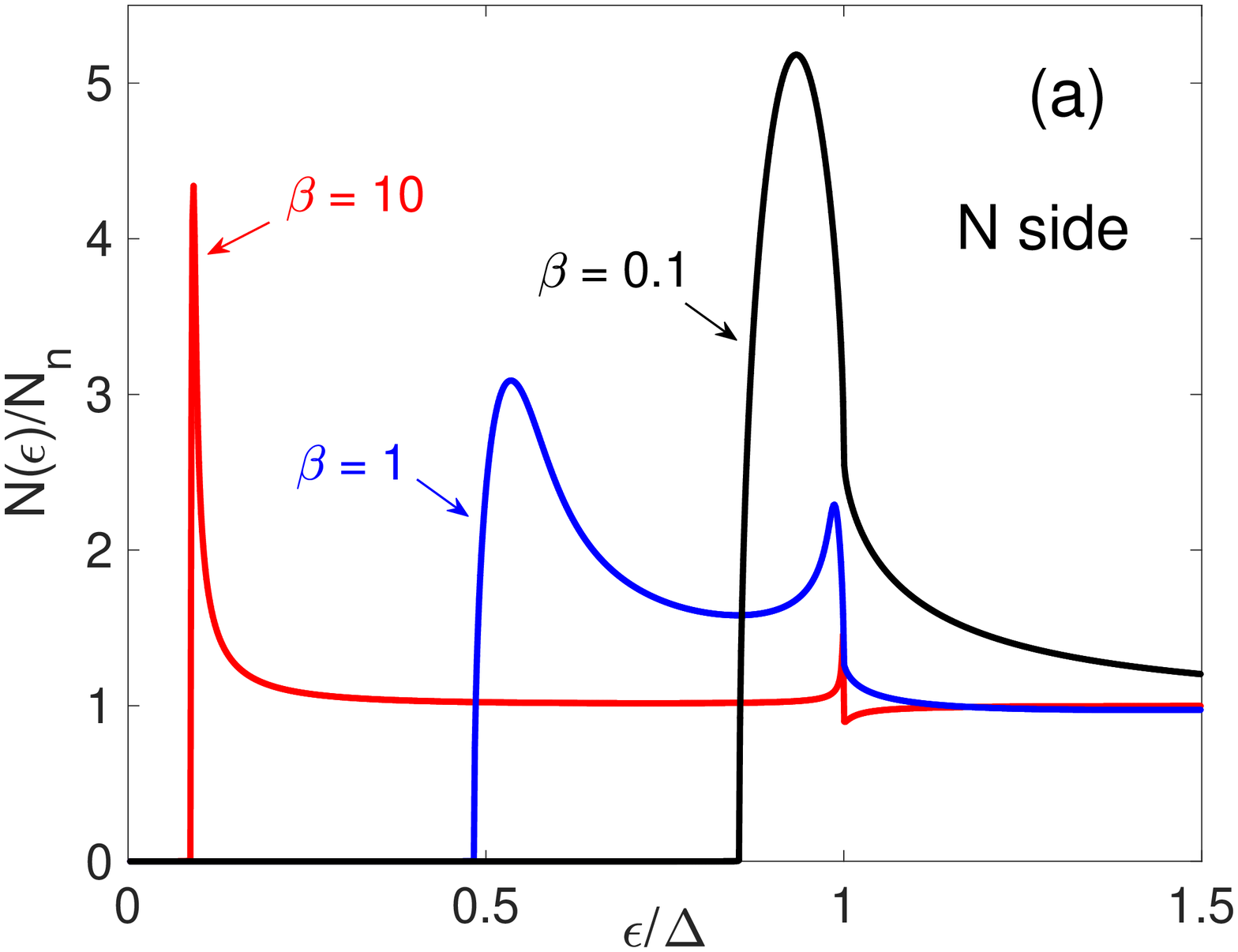}
    \\
    \vspace{-1mm}
   \includegraphics[width=9cm]{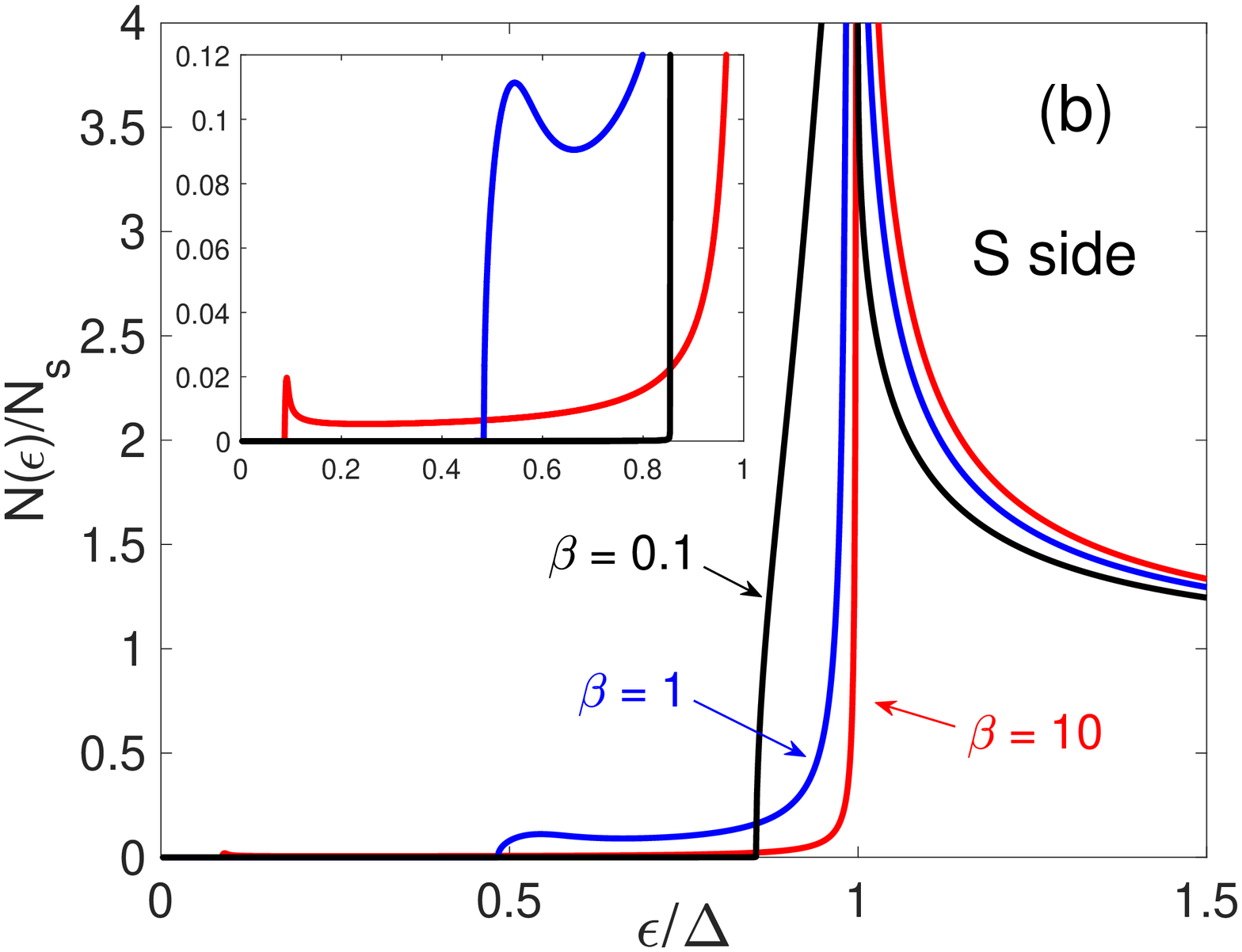}
   \caption{
Densities of states at (a) N side and (b) S side of the interface, 
calculated for $\alpha=0.05$, $\Gamma = 0$, $\hbar\Omega=11\Delta$, and $\beta =0.1,~1,~10$.
Inset in (b) shows details of the proximity-induced low-energy tail of $N(\epsilon)$ at S side.  
   }\label{fig4}
   \vspace{-3mm}
\end{figure}
Using Eqs.~(\ref{GR}) and (\ref{eq:NS_GR_Nx}), we obtain the density of states $n(\epsilon,x)=\mbox{Re}G^R(\epsilon,x)$: 
\begin{gather}
n_N(\epsilon, x) = {\rm Re}\biggl[  \frac{\cosh\theta_0-i\beta\epsilon}{\sqrt{1 -\beta^2\epsilon^2-2i\beta \epsilon \cosh\theta_0 }} \nonumber \\
- \frac{ i\epsilon x(x+2d)\sinh^2\theta_0}{2\xi_N^2[1 -\beta^2\epsilon^2-2i\beta \epsilon \cosh\theta_0]} \biggr] , 
\label{dos_N}\\
n_S(\epsilon, x)={\rm Re}\biggl[\frac{\epsilon(1+6t^2+t^4)+4t(1+t^2)}{(1-t^2)^2\sqrt{\epsilon^2-1}}\biggr], 
\label{dosSS}
\end{gather}
where $t(x)$ is defined by Eq. (\ref{t}). For $\epsilon$ not too close to $1$ so that $|\delta\theta_0(\epsilon)|\ll 1$, Eq. (\ref{dosSS}) simplifies to:
\begin{equation}
n_S(\epsilon, x) \simeq {\rm Re}\biggl[ \frac{\epsilon}{\sqrt{\epsilon^2-1}}+\frac{\delta\theta_0 e^{-k_{\epsilon}x}}{\sqrt{\epsilon^2-1}}  \biggr], \quad x>0. 
\label{dos_S}
\end{equation}
Figures~\ref{fig4}(a) and (b) show DOS at the N and S sides ($x= \mp 0$) of the interface, respectively. 
For a nearly-transparent interface with $\beta\ll 1$,  a thin N region disturbs DOS weakly so that $n(\epsilon)$ is close to the BCS DOS both in S and N regions which are coupled strongly by the proximity effect. As $\beta$ increases, N and S regions become more and more decoupled resulting in subgap states in the N region and the quasiparticle mini gap $\epsilon_0<1$ which decreases with $\beta$. By contrast, DOS at the S region approaches the BCS-like DOS as $\beta$ increases.  

\begin{figure}[tb]
   \begin{center}
   \includegraphics[width=1\linewidth]{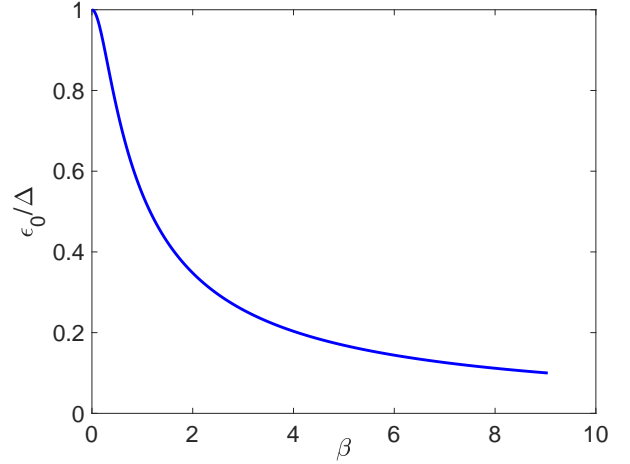}
   \end{center}\vspace{0cm}
   \caption{
   Minigap $\epsilon_0$ in the N layer as a function of $\beta$ calculated from Eq. (\ref{be}). 
   }
   \label{fig5}
\end{figure}

To see under what conditions the minigap in the N region can drop well below the bulk $\Delta$, we evaluate $\epsilon_0$ in the limit of $\Gamma=0$ and $\alpha \ll 1$ for which $\epsilon_0$ is an endpoint at which the density of states $N(\epsilon)$ vanishes. As follows from Eqs.~(\ref{eq:Phi_def}) and (\ref{eq:self_consistency_gamma2}), we have $\delta\theta_{0} \ll 1$ for $\alpha\ll 1$ and arbitrary $\beta$. 
In the zeroth order approximation in $\delta\theta_{0}$, the condition $n(\epsilon,-0) =0$ reduces to finding the root of the equation
$(1- \beta^2\epsilon_0^2)(1-\epsilon_0^2) =2\beta \epsilon_0^2 (1-\epsilon_0^2)^{1/2}$. This yields the following 
explicit dependence of $\beta$ on $\epsilon_0$:
\begin{equation}
\beta=\frac{1}{\epsilon_0}\left(\frac{1-\epsilon_0}{1+\epsilon_0}\right)^{1/2}.
\label{be}
\end{equation} 
As $\beta$ increases the minigap $\epsilon_0$ decreases as shown in Fig.~\ref{fig5}. 
The behavior of $\epsilon_0(\beta)$ in two limiting cases are:
\begin{eqnarray}
\epsilon_0\simeq 1-2\beta^2,\qquad \beta \ll 1,
\label{e01}\\
\epsilon_0 = \beta^{-1},\qquad \beta\gg 1. 
\label{e02}
\end{eqnarray}
Expressing Eq. (\ref{e02}) in dimensional units shows that the mini gap $\epsilon_0$ in a weakly-coupled layer $(\beta\gg 1)$ is independent of superconducting parameters:
\begin{equation}
\epsilon_0=\frac{\hbar}{4e^2N_ndR_B},\qquad \beta\gg 1.
\label{mgap}
\end{equation}

The N layer affects DOS in the S region as shown in Fig.~\ref{fig4}(b) where $n(\epsilon)$ calculated from Eq.~(\ref{dos_S}) at $\epsilon > \epsilon_0$ 
$\alpha=0.05$, $\Gamma=0$ and different values of $\beta$ are plotted. Insets show the  respective behaviors of $n(\epsilon)$ at small $\epsilon$ in a model with a finite $\Gamma$ independent of energy.


%
\begin{figure}[tb]
   \includegraphics[width=9cm]{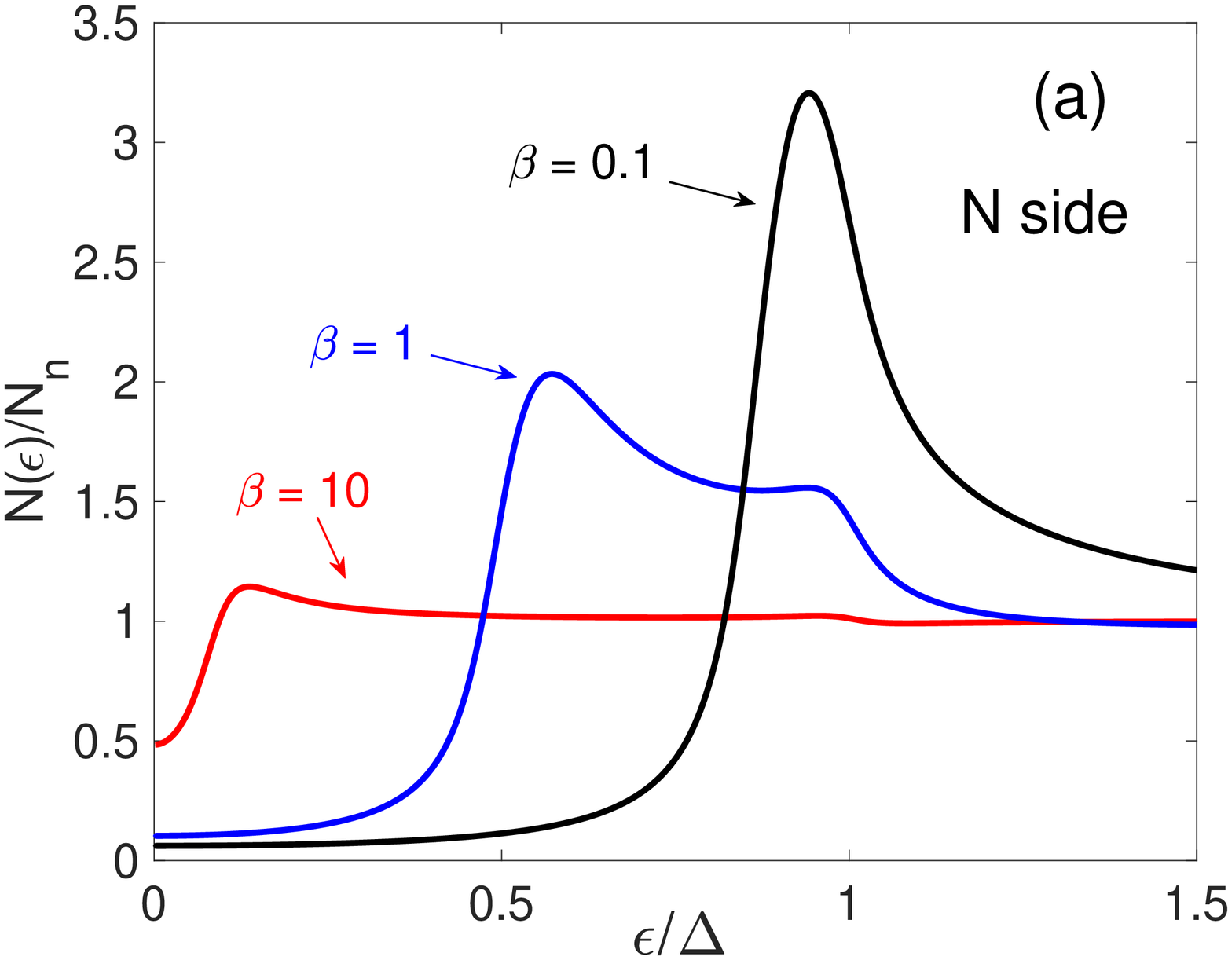}
   \\
    \includegraphics[width=9cm]{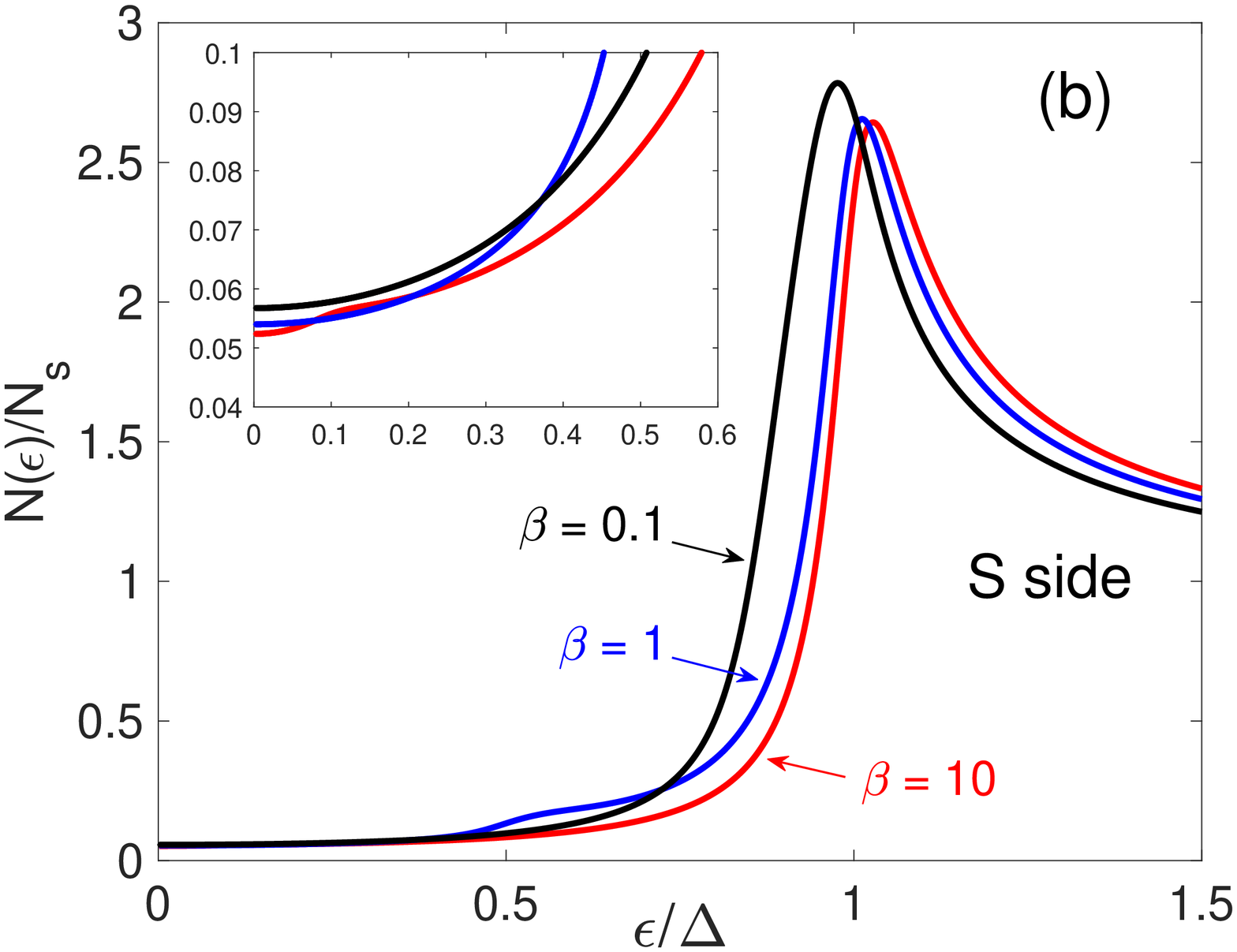}
   \caption{
Densities of states at (a) N side and (b) S side of the interface calculated for  
$\alpha=0.05$, $\Gamma = 0.05\Delta$, $\hbar\Omega=11\Delta$, and $\beta=0.1,~1,~10$. 
Inset in (b) shows $N(\epsilon)$ at $\epsilon\ll \Delta$. 
   }
   \label{fig6}
\end{figure}

The effect of a finite quasiparticle lifetime on DOS is calculated by replacing $\epsilon \to \epsilon + i\Gamma$ in Eqs. (\ref{dos_N}) and (\ref{dos_S}).
Taking $\Gamma$ into account smears out the cusps in Fig.~\ref{fig4} and causes a finite density of subgap states at $\epsilon =0$, as shown in Fig. 6. 
In the most interesting case of $\alpha \ll 1$, the zero-energy values of $n_N(0)$ and $n_S(0)$ at the N and S sides of the interface can be evaluate from Eq. (\ref{dos_N}) and (\ref{dos_S}) in the zero order approximation in $\delta\theta_0\ll 1$ at arbitrary transparency parameter $\beta$:
\begin{gather}
n_N(0) 
=\frac{\Gamma (1+\beta \sqrt{1+\Gamma^2})}{[(1+\beta^2\Gamma^2)(1+\Gamma^2) + 2\beta\Gamma^2\sqrt{1+\Gamma^2}]^{1/2}},
\label{eq:subgap_states_N}
\\
n_S(0) = \frac{\Gamma}{\sqrt{1+\Gamma^2}}. 
\label{eq:subgap_states_S}
\end{gather}
As $\beta$ increases, $n_N(0)$ approaches the normal density of states for a fully decoupled N layer at $\beta\gg 1$.

\section{Surface impedance}
Here we use the results of previous sections to calculate the effect of imperfect surface on the surface impedance $Z=R_s+iX$, where the reactive part $X=\mu_0\omega\lambda$ is expressed in terms of the global London penetration depth $\lambda$.  The impedance is calculated by expressing the complex conductivity $\sigma=\sigma_1-i\sigma_2$ in the current density $\textbf{J}=\sigma\textbf{E}$ in terms of retarded Green's functions \cite{kopnin,Belzig_review} as summarized in Appendix D.  The surface impedance $Z=E(0)/H(0)$ can be presented in an alternative form using the Fourier transform of the Maxwell equation $E'=-i\omega B$, so that $E(0)=i\omega\mu_0\int_0^\infty H(x)dx$, where $\omega$ is the frequency of the applied field $H(t)=H_ae^{i\omega t}$:
\begin{equation}
Z=\frac{i\mu_0\omega}{H_a}\int_0^\infty H(x)dx
\label{zz}
\end{equation}
Comparing Eq. (\ref{zz}) with $Z=i\omega\mu_0\lambda +R_s$, it is convenient to define the quasi-static global penetration depth $\lambda$ 
of a superconducting bilayer in terms of the in-phase component of the magnetic field $H(x)$, 
\begin{equation}
\lambda=\frac{1}{H_a}\int_0^\infty H(x)dx.
\label{tlam}
\end{equation}
In what follows we limit ourselves to the local London limit $\lambda\gg\xi_S$ in a dirty superconductor. A general case of dc magnetic screening in N-S bilayers with an arbitrary mean free path was addressed in Ref. \onlinecite{dc4}.

\subsection{Surface reactance and global penetration depth}

Using the temperature Green's functions, we calculate here the quasi-static penetration depth $\lambda$ for: 1. A superconductor with an ideal surface but with a finite $\Gamma$, 2. A superconductor with N surface layer and $\Gamma=0$. 

\subsubsection{Effect of bulk subgap states}
For a superconductor with an ideal surface but finite $\Gamma$, the penetration depth is given by:
\begin{equation}
\frac{1}{\lambda^2}=\frac{4\pi\mu_0 k_BT}{\hbar\rho_s}\sum_{\omega_n>0}\frac{\Delta^2}{(\omega_n+\Gamma)^2+\Delta^2},
\label{lamb}
\end{equation}  
where $\rho_s$ is the resistivity of a superconductor in the normal state. Equation (\ref{lamb}) is a generalization of the expression for $\lambda$ in the dirty limit \cite{kopnin} 
with the replacement $\omega_n\to \omega_n+\Gamma$. The sum in Eq. (\ref{lamb}) can be expressed in terms of a digamma function $\psi(z)$:
\begin{equation}
\frac{1}{\lambda^2}=\frac{2\mu_0\Delta}{\hbar\rho_s} {\rm Im}\psi\left(\frac{1}{2}+\frac{\Gamma}{2\pi k_B T}+\frac{i\Delta}{2\pi k_BT} \right). 
\label{lam}
\end{equation}
Since ${\rm Im}\psi(1/2 + ix)=(\pi/2)\tanh\pi x$, Eq. (\ref{lam}) at $\Gamma=0$ reproduces the well-known result \cite{kopnin}
\begin{equation}
\frac{1}{\lambda^2}=\frac{\pi\mu_0\Delta}{\hbar\rho_s}\tanh\frac{\Delta}{2 k_B T}.
\label{lam0}
\end{equation}
If $\Gamma > 0$ and $T \ll T_c$ we use the asymptotic expansion $\psi (z) = \ln z - 1/2z-1/12z^2$ and obtain:
\begin{equation}
\frac{1}{\lambda^2}=\frac{2\mu_0\Delta}{\hbar\rho_s}\left[\tan^{-1}\frac{\Delta}{\Gamma}-\frac{\pi^2k_B^2T^2\Gamma\Delta}{3(\Gamma^2+\Delta^2)^2} \right].
\label{lam1}
\end{equation}
At $\Gamma=0$ the penetration depth described by Eq. (\ref{lam0}) has the BCS exponential temperature dependence, $\lambda(T) - \lambda(0)\propto \exp (-\Delta/k_BT)$ at $k_BT\ll \Delta$. However, as follows from Eq. (\ref{lam1}), the subgap states in the Dynes model can change this BCS behavior of $\lambda(T)$, resulting instead in a quadratic temperature dependence of $\lambda(T)$ at $T\ll T_c$ in a s-wave superconductor. This may be essential for the interpretation of experimental data as observations of a power-law temperature dependence of $\lambda(T)$ has been usually attributed to a nodal pairing symmetry. 

\subsubsection{Effect of a normal surface layer.}

Let us define partial field penetration depths $\lambda_N$ and $\lambda_S$ in N and S regions:
\begin{gather}
\frac{1}{\lambda_N^2}=\frac{4\pi\mu_0 k_BT}{\hbar\rho_n}\sum_{\omega_n>0}\sin^2\theta_-,
\label{lamn} \\
\frac{1}{\lambda_S^2}=\frac{4\pi\mu_0 k_BT}{\hbar\rho_s}\sum_{\omega_n>0}\sin^2\theta_\infty,
\label{lams}
\end{gather}  
where $\theta_-$ is given by Eq. (\ref{sin-}). Here we assume that $\alpha\ll 1$ and neglect the small correction $\delta\theta(x)$ in S region.
Next we solve the London equations $\lambda_N^2H_N''=H_N$ and $\lambda_S^2H_S''=H_S$ in N and S regions with the boundary conditions, $H_N(0)=H_a$, $H_N(d)=H_S(d)$, and $\lambda_S^2H_S'(d)=\lambda_N^2H_N'(d)$ which yields \cite{kubo,aml}: 
\begin{gather}
H_N=H_a[(1-c)e^{-x/\lambda_N}+ce^{x/\lambda_N}], \quad 0<x<d,
\label{Hn}\\
H_S=H_abe^{(d-x)/\lambda_S}, \qquad x>d,
\label{Hs}
\end{gather}
where $c=k/(k+e^{2d/\lambda_N})$, $b=(1+k)/(ke^{-d/\lambda_N}+e^{d/\lambda_N})$, and $k=(\lambda_N-\lambda_S)/(\lambda_N+\lambda_S)$. Using Eqs. (\ref{tlam}), (\ref{Hn}) and (\ref{Hs}), we calculate the global penetration depth: 
\begin{equation}
\lambda=\frac{(e^{d/\lambda_N}-1)(e^{d/\lambda_N}+k)\lambda_N}{k+e^{2d/\lambda_N}}+\frac{(1+k)e^{d/\lambda_N}\lambda_S}{k+e^{2d/\lambda_N}}.
\label{lame}
\end{equation}
Eq. (\ref{lame}) yields the obvious limits $\lambda=\lambda_N$ if $d\gg\lambda_N$, and $\lambda=\lambda_S$ if $\lambda_N=\lambda_S$ or $d\to 0$. Here we are interested in the case of $d\ll \lambda_N$ where $\lambda$ is close to $\lambda_S$ and the surface layer results in a small correction which is calculated by expanding  Eq. (\ref{lame}) in the first order in $d$:
\begin{equation}
\lambda=\lambda_S+d\left(1-\frac{\lambda_S^2}{\lambda_N^2}\right).
\label{laml}
\end{equation}
Here the temperature dependence of the ratio $\lambda_S^2/\lambda_N^2$ can be determined by the small mini gap $\epsilon_0$ if the interface transparency parameter $\beta $ is large.   We illustrate this effect in the weak transparency limit of $\beta\gg 1$ at $\Gamma=0$ for which the $\omega$-summation in Eqs. (\ref{lamn}) and (\ref{lams}) can be done exactly using Eq. (\ref{sin-}) where the term $\beta\omega\cos\theta_0$ in the denominator can be neglected. Hence,
\begin{equation}
\lambda=\lambda_S+d-\frac{d\epsilon_0\sigma_n}{\Delta\sigma_s}\left[\frac{\tanh(\epsilon_0/2k_BT)}{\tanh(\Delta/2k_BT)}-\frac{\epsilon_0}{\Delta} \right],
\end{equation}
where the mini gap $\epsilon_0$ is given by Eq. (\ref{mgap}). The first term in the brackets gives the main temperature dependence at $k_B T \ll \Delta$ where the BCS contribution $\propto\exp(-\Delta/k_BT)$ becomes negligible.  If $\beta\gg 1$, we have $\theta_-\ll\theta_\infty$, so that $\lambda_N\gg\lambda_S$, and $\lambda\to\lambda_s+d$ because the decoupled N layer provides no screening. Calculations of $\lambda(T)$ in a SN bilayer model at $\Gamma=0$ were performed in Ref. \onlinecite{lamn}.
 
\subsection{Surface resistance}
In the case of $d\ll \xi_S \ll \lambda$ and  $\hbar\omega\ll\Delta$ considered in this work, the RF field is constant in the N region ($-d < x <0$) and attenuates at $x \ge 0$, so   
the amplitude of a low-frequency vector potential $A(x,t)=A(x)e^{i\omega t}$ follows the quasi-static Meissner distribution: 
\begin{eqnarray}
A(x) =
\begin{cases}
-\lambda B_0 & (x <0), \\
-\lambda B_0 e^{-\frac{x}{\lambda}} & (x\ge 0).
\end{cases}  
\end{eqnarray}
The Fourier components of electric field and the current density are then $E(x) =-i\omega A(x)$ 
and $J(x) = -i\omega \sigma(x) A(x)$, respectively, where $\sigma=\sigma_1-i\sigma_2$ is the complex conductivity.  The surface resistance $R_s$ can be expressed in terms of $\sigma(\omega)$ using 
the power generated by the $RF$ currents per unit surface, $(1/2) R_s H_0^2 = (1/2) \int {\rm Re} [E J^*] dx= (1/2)\omega^2 \int \sigma_1 A^2dx$. Hence,
\begin{eqnarray}
R_s = \omega^2\mu_0^2\lambda^2 
\biggl[ \int_{-d}^0 \!\! dx \sigma_1^{\rm N}(x) + \int_0^{\infty} \!\!\!dx\sigma_1^{\rm S}(x) e^{-\frac{2x}{\lambda}} \biggr] . \label{eq:general_Rs}
\end{eqnarray}
Here the local dissipative conductivities $\sigma_1^{\rm N,S}(\omega,x)$ are expressed in terms of the respective Green's functions as follows (see Appendix D):
\begin{gather}
\sigma_1^i(\omega) =\frac{\sigma_{n,s}}{\hbar\omega} \int_{-\infty}^{\infty} \!\!\!d\epsilon [ f(\epsilon) - f(\epsilon + \hbar\omega)]\times \nonumber \\
[ n_i(\epsilon,x) n_i(\epsilon + \omega,x) + m_i(\epsilon, x) m_i(\epsilon + \omega,x) ],  \label{sigg}
\end{gather}
where $i=N,S$, $f(\epsilon)=(e^{\epsilon/k_BT}+1)^{-1}$, $n(\epsilon,x)={\rm Re} \cosh\theta(\epsilon,x)$, and $m(\epsilon,x)={\rm Re} \sinh\theta(\epsilon,x)$.
Using Eqs. (\ref{eq:general_Rs}) and (\ref{sigg}), we present $R_s$ in the form
\begin{gather}
R_s = \frac{\omega}{\hbar}\left(e^{\hbar\omega/k_BT}-1\right)\mu_0^2\lambda^2( \sigma_nI_N + \sigma_sI_S) ,
\label{eq:general_Rs_2} \\
\!\!I_N = d\int_{-\infty}^{\infty} \frac{\cosh(u^N+u^N_+)\cos v^N\cos v^N_+}{[1+e^{-\epsilon/k_BT}][e^{(\epsilon+\hbar\omega)/k_BT}+1]} d\epsilon,
 \label{eq:IN} \\
\!\!\!\!I_S =\!\int_0^\infty\!\!dx e^{-2x/\lambda}\!\!\int_{-\infty}^{\infty}\!\frac{d\epsilon \cosh (u^S+u^S_+)\cos v^S\cos v^S_+}{[1+e^{-\epsilon/k_BT}][e^{(\epsilon+\hbar\omega)/k_BT}+1]}. 
\label{eq:IS}
\end{gather}
Here $u(\epsilon,x)$ and $v(\epsilon,x)$ are defined by the real-frequency solutions of the Usadel equation, $\theta(\epsilon,x)=u(\epsilon,x) + iv(\epsilon,x)$, $u_+=u(\epsilon+\hbar\omega,x)$, $v_+=v(\epsilon+\hbar\omega,x)$, and the factors $I_N$ and $I_S$ represent contributions from N and S regions, respectively.  In Eq. (\ref{eq:IN}) we neglect a weak variation of $\theta(x)$ across a thin N layer, assuming that $\theta^N=\theta_-$ at $\alpha\ll 1$, where $\theta_-$ is defined by Eq. (\ref{eq:NS_FR_N}).

Equations (\ref{eq:general_Rs_2})-(\ref{eq:IS}) determine the surface resistance at $\hbar\omega <\Delta$, taking into account the significant variation of local DOS due to proximity and pairbreaking effects at the surface considered above. Here $R_s$ turns out to be quite sensitive to low-energy tails of DOS both in N and S regions, so Eqs. (\ref{eq:general_Rs_2})-(\ref{eq:IS}) should be solved numerically together with Eqs. (\ref{GR})-(\ref{t}), (\ref{eq:NS_real_theta0}) and (\ref{dosSS}) which describe self-consistently the relevant superconductring properties at a non-ideal interface. In the following, we calculate $R_s$ for several characteristic cases. 

\subsubsection{Ideal surface with bulk subgap states}

For an ideal surface with no pairbreaking layers $(I_N=0)$ but subgap states in the bulk $(\Gamma>0)$, Eqs. (\ref{eq:general_Rs})-(\ref{eq:IS}) in the dimensional units become:
\begin{gather}
R_s= \frac{\omega}{2\hbar}\int_{-\infty}^{\infty}\frac{(e^{\hbar\omega/k_BT}-1)\mu_0^2\lambda^3\sigma_s d\epsilon}{(1+e^{-\epsilon/k_BT})[e^{(\epsilon+\hbar\omega)/k_BT}+1]}  
\nonumber \\
\times [ n(\epsilon) n(\epsilon + \hbar\omega) + m(\epsilon) m(\epsilon + \hbar\omega)],
\label{Rss} \\
n(\epsilon)={\rm Re}\frac{\tilde{\epsilon}}{\sqrt{\tilde{\epsilon}^2-\Delta^2}}, \quad
 m(\epsilon)={\rm Re}\frac{\Delta}{\sqrt{\tilde{\epsilon}^2-\Delta^2}},
\label{nm}
\end{gather}
where $\tilde{\epsilon}=\epsilon+i\Gamma$. Consider first the case of $\Gamma=0$ and $\hbar\omega\ll\Delta$ in which the integration in Eq. (\ref{Rss}) can be restricted to $\Delta<\epsilon<\infty$, the regions of negative and positive $\epsilon$ giving equal contributions. At $T\ll T_c$ the main contribution to the integral in Eq.~(\ref{Rss}) 
comes from a narrow range of energies,  
$\epsilon -\Delta \sim k_B T \ll \Delta$, so that the denominators of $n(\epsilon)$ and $m(\epsilon)$ can be replaced with $\sqrt{2\Delta(\epsilon - \Delta)}$, and the integration yields \cite{Zmuidzinas}:
\begin{equation}
\!R_s=\frac{2\mu_0^2\omega\lambda^3\Delta}{\hbar\rho_s}\sinh\!\left[\frac{\hbar\omega}{2k_BT}\right]\!K_0\!\left[\frac{\hbar\omega}{2k_BT}\right]e^{-\Delta/k_BT},
\label{Rs0}
\end{equation} 
where $\rho_s=1/\sigma_s$ is the normal state resistivity, and $K_0(x)$ is a modified Bessel function. At low frequencies $\hbar\omega \ll k_BT$ such as 
$\omega/2\pi \sim 1\,{\rm GHz}$ and $T\simeq 2\,{\rm K}$, we have $\hbar \omega/2 k_B T \sim 10^{-2}$ so that Eq. (\ref{Rs0}) simplifies to \cite{ag_sust}:
\begin{equation}
R_s\simeq \frac{\mu_0^2\omega^2\lambda^3\Delta}{\rho_sk_BT}\ln\left[\frac{C_1k_BT}{\hbar\omega} \right]e^{-\Delta/k_BT},
\label{Rsa}
\end{equation}
where $C_1=4e^{-\gamma_E}\approx 9/2$. The factor $\ln(k_BT/\hbar\omega)$ in Eq. (\ref{Rsa}) results from a logarithmic singularity 
in $\sigma_1(\omega)$ at $\hbar\omega\to 0$ as two square root singularities in the integrand of Eq. (\ref{sigg}) merge into a pole. This feature of DOS in the 
idealized BCS model disappears if the realistic broadening of the gap singularities in $N(\epsilon)$ is taken into account, resulting in a finite $\sigma_1(\omega)$ at $\omega=0$.

Eq. (\ref{Rss}) describes both the exponential BCS contribution $R_s \propto \exp(-\Delta/k_BT)$ and an additional  term $R_i(T)$ that is not exponentially small at $T\ll T_c$. Here $R_i$ can be evaluated at $\Gamma\gg\hbar\omega$, and $k_BT\ll \Gamma$ for which $f(\epsilon)-f(\epsilon+\hbar\omega)$ in Eq. (\ref{sigg}) is a sharp peak of width $k_BT$ at $\epsilon=0$, so $n(\epsilon)$ and $m(\epsilon)$ can be expanded up to quadratic terms in $\epsilon\ll k_BT$. Then Eq. (\ref{sigg}) yields a finite conductivity, resulting in a residual resistance in Eq. (\ref{Rss}) at $k_BT \lesssim \Gamma$:
\begin{equation}
R_i(T)=\frac{\mu_0^2\omega^2\lambda^3 \Gamma^2}{2\rho_n(\Delta^2+\Gamma^2)}\biggl[1+\frac{4\pi^2k_B^2T^2\Delta^2}{3(\Delta^2+\Gamma^2)^2}\biggr],
\label{rii} 
\end{equation}
where $\Delta(\Gamma)$ is given by Eq. (\ref{delg}), and the temperature-dependent correction in the brackets was obtained for $\hbar\omega\ll k_BT$. From Eq. (\ref{rii}), it follows that $R_i\simeq 10$ n$\Omega$ observed on large-grain Nb cavities at $1.5$ GHz ~\cite{clare,padams}, $\lambda = 40$ nm, and $\rho_n=1$ n$\Omega\cdot$m corresponds to $\Gamma \simeq 0.05\Delta$.

\begin{figure}[tb]
\begin{center}
\includegraphics[width=9.8cm]{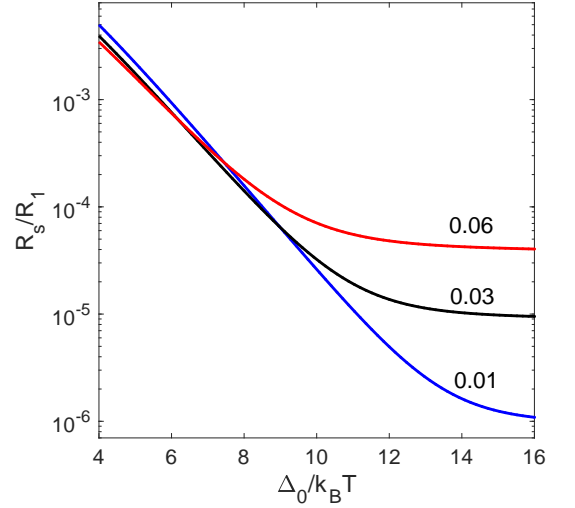}
\end{center}
\vspace{-3mm}
\caption{Arrhenius plots for $R_s(T)$ calculated from Eq. (\ref{Rss})-(\ref{nm}) for $\hbar\omega=0.01\Delta$ and $\Gamma/\Delta_0=0.01$, $0.03$ and $0.06$. 
Here $R_1=\mu_0^2\lambda^3\omega\Delta/2\hbar\rho_s$, the temperature dependencies of $\Delta$ and $\lambda$ at $T<T_c/2$ are neglected, and $\Delta(\Gamma)$ is given by 
Eq. (\ref{delg}) }
\label{ri}
\end{figure}

Shown in Fig. \ref{ri} is the Arrhenius plot of $R_s(T)$ calculated from Eqs. (\ref{Rss})-(\ref{nm}) for different ratios of $\Gamma_0/\Delta$. Here $\ln[R_s(T)]$ at higher $T$ follows the linear dependence on $1/T$ expected from the BCS model,  but at lower temperatures $\ln[R_s(T)]$ levels off, which would be usually attributed to a residual resistance. Here $R_i$ is a part of the BCS surface resistance including a realistic broadening of the gap peaks in $N(\epsilon)$. Moreover, Fig. \ref{ri} shows that increasing $\Gamma$ {\it reduces} $R_s(T)$ at higher temperatures for which the residual resistance is negligible.

The finite $R_i$ in Eq. (\ref{rii}) results from a nonzero DOS at the Fermi level in the Dynes model, while a reduction $R_s$ with $\Gamma$ at higher $T$ comes from the reduction of the logarithmic factor in Eq. (\ref{Rsa}). If $\Gamma>0$, the square root gap singularities in $n(\epsilon)$ and $m(\epsilon)$ in Eq. (\ref{nm}) turn into finite peaks of width $\sim\Gamma$.  At $\Gamma >\hbar\omega$ but $\Gamma\ll k_BT$ integration of these peaks in Eq. (\ref{Rss}) at $\epsilon\simeq\Delta$ yields a logarithmic term similar to that of in Eq. (\ref{Rsa}) but with energy cutoff $\Gamma$ instead of $\hbar\omega$.  Therefore, the smearing of the gap peaks in $N(\epsilon)$ on $R_s$ can be qualitatively taken into account by the replacement: $\ln (k_BT/\hbar\omega) \to \ln (k_BT/\Gamma)$.    
Such broadening of the DOS peaks reduces $R_s(T)$ at temperatures $T\gg \hbar\omega/k_B$ at which $R_i$ is negligible, as it is clearly seen in Fig. \ref{ri}. Moreover, $R_s(T)$ can also be reduced by pairbreaking mechanisms which suppress superconductivity, as will be shown below.  

\subsubsection{Ideal surface with paramagnetic impurities in the bulk}

\begin{figure}[tb]
\includegraphics[width=9cm]{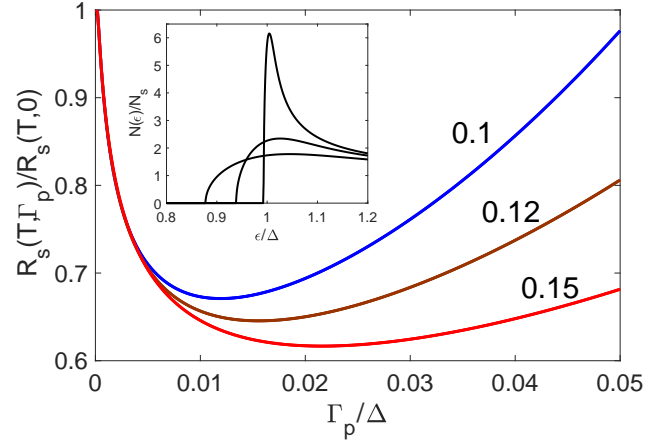}
\caption{Minimum in the surface resistance $R_s(\Gamma_p)$ as a function of the spin-flip pairbreaking parameter $\Gamma_p$ calculated from Eqs. (\ref{gaps})-(\ref{Rsp}) at $\hbar\omega=0.005\Delta$ and $k_BT/\Delta=0.1,~ 0.12,~ 0.15$. Inset shows $N(\epsilon)=N_s\cosh u\cos v$ calculated from Eqs. (\ref{v})-(\ref{r}) at $\Gamma_p/\Delta= 0.001,~0.02,~0.05$. }
\label{rsmagg}
\end{figure}

It is well-known that spin-flip pairbreaking scattering on paramagnetic impurities broadens the peaks in $N(\epsilon)$ and reduces the quasiparticle gap~\cite{maki}:
\begin{equation}
\epsilon_g=\left(\tilde{\Delta}^{2/3}-\Gamma_p^{2/3}\right)^{3/2}.
\label{gaps}
\end{equation}  
Here $\Gamma_p= 4\pi n_pN_sS(S+1)J^2$ is the spin-flip pairbreaking parameter in the Born approximation, where $n_p$ is the volume density of 
paramagnetic impurities with spin $S$, and $J$ is the exchange integral.  
The quasiparticle gap $\epsilon_g$ in Eq. (\ref{gaps}) is smaller than the order parameter $\tilde{\Delta}$ given by \cite{maki}
\begin{equation}
\tilde{\Delta} = \Delta-\frac{\pi}{4}\Gamma_p, \qquad \Gamma_p\ll\Delta.
\label{delts}
\end{equation}    
Here $\Delta$ is the order parameter in the absence of paramagnetic impurities. The behavior of $N(\epsilon)$ at different values of $\Gamma_p$ is shown in 
the inset of Fig. \ref{rsmagg}.
 
To calculate $R_s$ and the factors $n(\epsilon)=\mbox{Re}\cosh\theta$ and $m(\epsilon)=\mbox{Re}\sinh\theta$ in Eq. (\ref{eq:general_Rs_2}), we solve the uniform Usadel equation 
which takes into account spin-flip scattering on magnetic impurities:
\begin{equation}
\epsilon\sinh\theta+i\Gamma_p\cosh\theta\sinh\theta=\tilde{\Delta}\cosh\theta.
\label{usas}
\end{equation}
Writing $\theta=u+iv$, we find that the imaginary parts of Eq. (\ref{usas}) yields a quadratic equation for $\sin v$ which allows us to express $v$ in terms of $u$:
\begin{equation}
\sin v=\bigl[-\tilde{\Delta} +(\tilde{\Delta} ^{2}-\Gamma_p^{2}\sinh ^{2}2u)^{1/2}\bigr]/2\Gamma_p\cosh u.
\label{v}
\end{equation}
The real part of Eq. (\ref{usas}) yields the cubic equation $\Gamma_p^2\sinh ^{3}2u+(\epsilon ^{2}-\tilde{\Delta}^2 +\Gamma_p^{2})\sinh 2u-2\epsilon \tilde{\Delta}=0$ 
which has the following Cardano solution \cite{agprl}:
\begin{gather}
\sinh 2u =[(r+\epsilon \tilde{\Delta} \Gamma_p) ^{1/3}-(r-\epsilon \tilde{\Delta}\Gamma_p )^{1/3}]/\Gamma_p, 
\label{card}\\
r=[\epsilon ^{2}\tilde{\Delta} ^{2}\Gamma_p^{2}+
(\epsilon ^{2}+\Gamma_p^{2}-\tilde{\Delta} ^{2})^{3}/27]^{1/2},
\label{r} 
\end{gather}
Equations (\ref{v})-(\ref{r}) which define the spectral density $\cosh(u +u_+)\cos v\cos v_+$ in Eq. (\ref{eq:general_Rs_2}) are supplemented by 
the self-consistency equation for $\tilde{\Delta}$ which reduces to Eq. (\ref{delts}) at $\Gamma_p\ll\Delta$. 
The spectral density vanishes at $-\epsilon_g+\hbar\omega<\epsilon<\epsilon_g$, so the integration in Eq. (\ref{eq:general_Rs_2}) can be restricted to $\epsilon_g<\epsilon <\infty$, the regions of negative and positive $\epsilon$ giving equal contributions. If $\hbar\omega \ll k_BT$ and $\exp(-\Delta/k_BT)\ll 1$  the surface resistance is then
\begin{equation}
R_s = R_0
\int_{\epsilon_g}^{\infty}d\epsilon e^{-\epsilon/k_BT}
\cosh(u+u_+)\cos v\cos v_+,
\label{Rsp}
\end{equation}   
where $R_0=\omega^2\mu_0^2\lambda^3/2\rho_sk_BT$, $u_+=u(\epsilon+\hbar\omega)$, and $v_+=v(\epsilon+\hbar\omega)$.
Shown in Fig. \ref{rsmagg} is $R_s(\Gamma_p)$ as a function of the pairbreaking parameter $\Gamma_p$ calculated from Eqs. (\ref{delts})-(\ref{Rsp}) at different temperatures. 
There is a clear minimum in $R_s(\Gamma_p)$ which results from a competition of the broadening of the DOS peaks which reduces $R_s$ as $\Gamma_p$ increases, and the 
reduction of the quasiparticle gap $\epsilon_g$ which increases $R_s$ with $\Gamma_p$. The position of the minimum in $R_s(\Gamma_p)$ shifts to larger $\Gamma_p$ as
the temperature increases. These results show that a small density of magnetic impurities can noticeably (by $\sim 30-40\%$) decrease the surface resistance at low temperatures. To evaluate the 
conditions under which the minima in $R_s(\Gamma_p)$ occur, we notice that in the Abrikosov-Gor'kov theory of weak magnetic scattering used here, $T_c$ vanishes at 
$\Gamma_p =\hbar/2\tau_s=\Delta_0/2$, that is, $l_s\sim \xi_0$ where $\tau_s$ and $l_s$ are the spin flip scattering time and the mean free path, respectively, and $\xi_0=\hbar v_F/\pi\Delta_0$ 
is the clean limit coherence length at $T=0$. The values of $\Gamma_p \simeq (0.01-0.02)\Delta$ in Fig. 8 thus correspond to $l_s\sim 10^2\xi_0$, assuming no low-energy bound states on magnetic 
impurities \cite{balatski}.

\subsubsection{Reduced BCS coupling constant at the surface}

The surface resistance is given by  
\begin{gather}
R_{s}= \int_0^\infty\! dx \int_{-\infty}^{\infty}\!\frac{\omega (e^{\hbar\omega/k_BT}-1)\mu_0^2\lambda^2\sigma_s d\epsilon}{(1+e^{-\epsilon/k_BT})[e^{(\epsilon+\hbar\omega)/k_BT}+1]}\times
\nonumber \\
[ n(\epsilon,x) n(\epsilon + \hbar\omega,x) + m(\epsilon,x) m(\epsilon + \hbar\omega,x)]e^{-2x/\lambda}
\label{Rsb} 
\end{gather}
Here $n(\epsilon,x)$ and $m(\epsilon,x)$ are obtained using Eqs. (\ref{GR}-(\ref{t}): 
\begin{gather}
n(\epsilon, x)={\rm Re}\biggl[\frac{\tilde{\epsilon}(1+6t^2+t^4)+4t(1+t^2)\Delta}{(1-t^2)^2\sqrt{\tilde{\epsilon}^2-\Delta^2}}\biggr], 
\label{nb} \\
m(\epsilon,x)={\rm Re}\biggl[\frac{(1+6t^2+t^4)\Delta+4\tilde{\epsilon}t(1+t^2)}{(1-t^2)^2\sqrt{\tilde{\epsilon}^2-\Delta^2}}\biggr], 
\label{mb} \\
t(x)=\tanh\left(\frac{\theta_0-\theta_\infty}{4}\right)e^{-k_\epsilon x}, \qquad\tilde{\epsilon}=\epsilon+i\Gamma,
\label{tt} 
\end{gather}
where $\theta_0$ is a solution of the self-consistency Eq. (\ref{eq:reducedBCS_theta0}) with $k_\epsilon=[\Delta^2-\tilde{\epsilon}^2]^{1/4}(2/\hbar D_s)^{1/2}$.

\begin{figure}
\includegraphics[width=9cm]{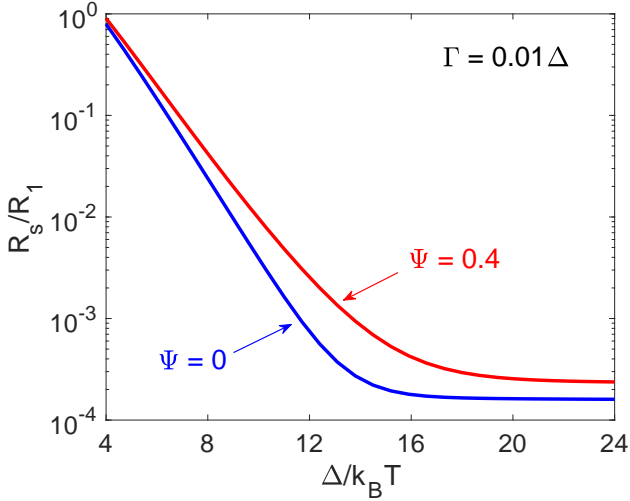}
\caption{$R_s(T)$ calculated from Eqs. (\ref{Rsb})-(\ref{tt}) and (\ref{eq:reducedBCS_theta0}) for $\Gamma=0.01\Delta$, $\lambda=5\xi_S$, and $\Psi = 0, ~0.4$. Here 
$R_1=\omega^2\mu_0^2\lambda^3\sigma_s/2$.}
\label{fig9}
\end{figure}

Shown in Fig. \ref{fig9} are the Arrhenius plots of $\ln R_s(T)$ versus $1/T$ calculated from Eqs. (\ref{Rsb})-(\ref{tt}) and (\ref{eq:reducedBCS_theta0}) for different values of the 
surface pairbreaking parameter $\Psi$. At $\Psi\ll 1$ the curves $R_s(T)$ reproduce the residual resistance determined by bulk subgap 
states which was calculated in the previous subsection. As $\Psi$ increases, the apparent high-temperature slope of $\ln R_s(T)$ (commonly used to extract the gap parameter $\Delta$ from the experimental data) decreases, 
and the plot of $\ln R(T)$ becomes curved even before it reaches the residual resistance plateau, which is also increases with $\Psi$ (by the factor $\sim 2$ for the case shown in Fig. \ref{fig9}). 
These features of $R_s(T)$ are manifestations of the 
broadening of the gap peaks in $N(\epsilon)$ at the surface shown in Fig. 2.     

\subsubsection{Thin normal layer}

For a superconductor with a thin N layer, the surface resistance can be written in the form:
\begin{equation}
R_s = \delta R + R_{s0},
\label{Rsnorm}
\end{equation}
where the bulk contribution $R_{s0}$ is given by Eqs. (\ref{Rsb})-(\ref{tt}), and the contribution from the N layer is:
\begin{gather}
\delta R = \frac{d}{\hbar} \int_{-\infty}^{\infty}\!\frac{\omega (e^{\hbar\omega/k_BT}-1)\mu_0^2\lambda^2\sigma_n }{(1+e^{-\epsilon/k_BT})[e^{(\epsilon+\hbar\omega)/k_BT}+1]}\times
\nonumber \\
[ n_N(\epsilon) n_N(\epsilon + \hbar\omega) + m_N(\epsilon) m_N(\epsilon + \hbar\omega)]d\epsilon,
\label{delR} \\
n_N(\epsilon) = {\rm Re}\biggl[\frac{\Delta\cosh\theta_0-i\beta\tilde{\epsilon}}{\sqrt{\Delta^2 -\beta^2\tilde{\epsilon}^2-2i\beta\tilde{\epsilon}\Delta \cosh\theta_0}}\biggr], 
\label{nN} \\
m_N(\epsilon)= {\rm Re} \biggl[ \frac{\Delta\sinh\theta_0}{\sqrt{\Delta^2 -\beta^2\tilde{\epsilon}^2-2i\beta\tilde{\epsilon}\Delta \cosh\theta_0}}\biggr], 
\label{mN} 
\end{gather}
where $\theta_0$ satisfies Eq. (\ref{eq:NS_real_theta0}) which takes into account pairbreaking and proximity effects 
(we assumed no magnetic impurities in the N layer~\cite{minigap_mag}). The ratio $\delta R/R_{s0}$ at $\beta=0$ is determined by the parameter,
\begin{equation}
\tilde{\alpha}=\frac{2d\sigma_n}{\sigma_s\lambda}=2\alpha\frac{D_n\xi_S}{D_s\lambda}
\label{alps}
\end{equation}

We evaluate the contribution of N layer to the residual resistance in the same way we did to derive Eq. (\ref{rii}).
At $k_B T \ll \Gamma$, the factor $f(\epsilon) - f(\epsilon+u)$ in Eqs.~(\ref{eq:IN}) and (\ref{eq:IS}) has a sharp peak with width $k_B T$ at $\epsilon = 0$, so that 
$n(\epsilon,x)$ and $m(\epsilon,x)$ in Eqs. (\ref{Rsb})-(\ref{mN}) can be replaced by their respective values at $\epsilon=0$: 
\begin{gather}
n_N(0) = \frac{\Gamma (1+ \beta\sqrt{1+\Gamma^2})}{\sqrt{(1+\beta^2\Gamma^2)(1+\Gamma^2)+2\beta\Gamma^2\sqrt{1+\Gamma^2}}}, \\
n_S(0) = \frac{\Gamma}{\sqrt{1+\Gamma^2}}, \qquad m_S(0) = m_N(0) =0. 
\end{gather}
Hence, we obtain in dimensional units:
\begin{gather}
R_i = \frac{1}{2}\mu_0^2 \omega^2 \lambda^3 \sigma_s\biggr[\frac{\Gamma^2}{\Delta^2+\Gamma^2} + \nonumber \\
\frac{\tilde{\alpha}\Gamma^2 \bigl( \Delta+ \beta\sqrt{\Delta^2+\Gamma^2} \bigr)^2}{(\Delta^2+\beta^2\Gamma^2)(\Delta^2+\Gamma^2)+2\beta\Gamma^2\Delta\sqrt{\Delta^2+\Gamma^2}}  \biggr] . 
\label{eq:subgap_Ri}
\end{gather}
If $d\to 0$, the second term in the brackets vanishes and Eq.~(\ref{eq:subgap_Ri}) reduces to Eq. (\ref{rii}) in which $R_i$ is determined by the bulk $\Gamma$. However,  
for a very high interface barrier $\beta\gg \Delta/\Gamma$, the parameter $\Gamma\ll \Delta$ in the second term in the bracket cancels out, and Eq. (\ref{eq:subgap_Ri}) yields the 
surface resistance $R_i=\mu_0^2\omega^2\lambda^2\sigma_nd+\mu_0^2\omega^2\sigma_s\lambda^3\Gamma^2/2\Delta^2$ of a decoupled N layer plus the bulk subgap contribution.

\begin{figure}
   \includegraphics[scale=0.4]{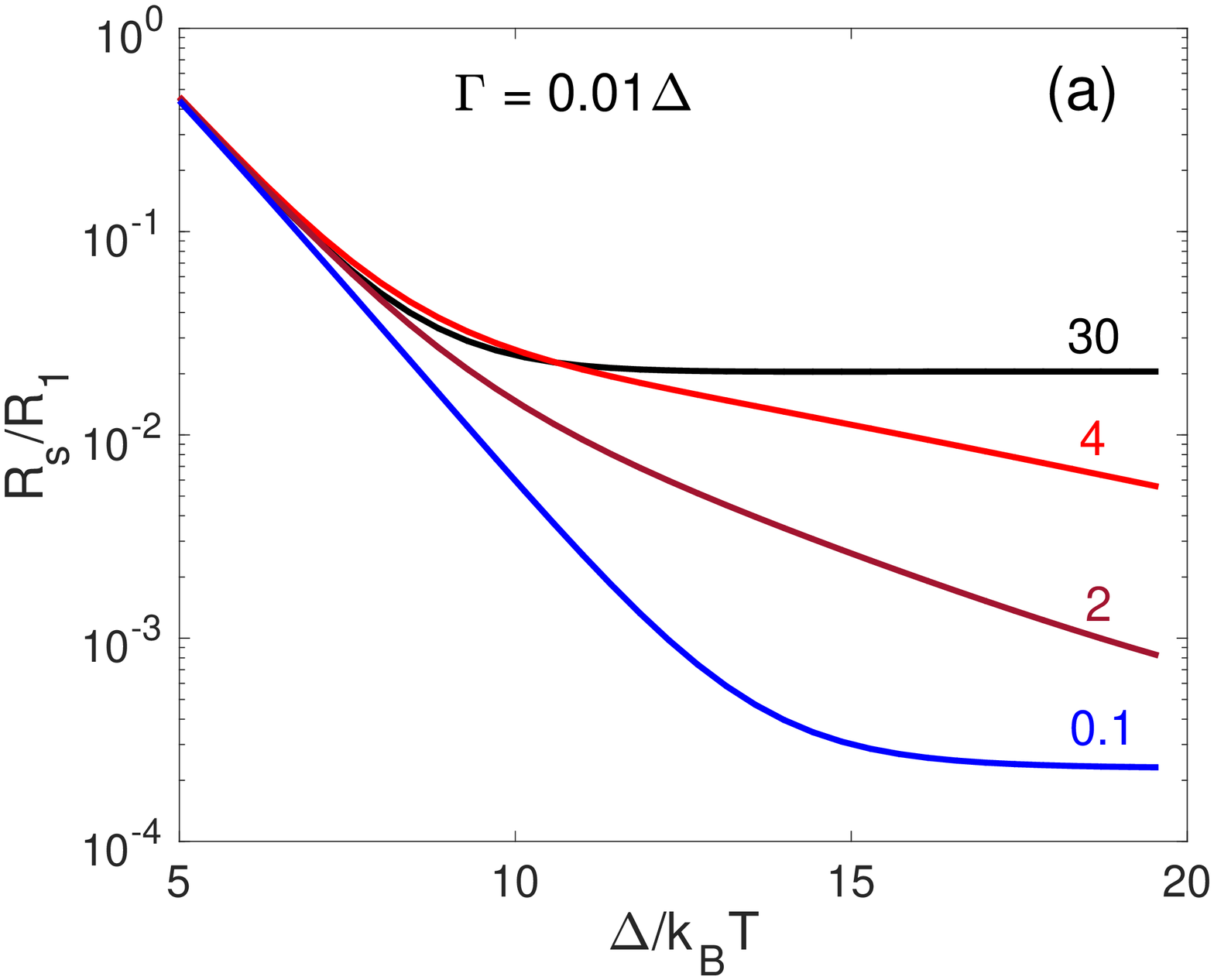}
   \includegraphics[scale=0.4]{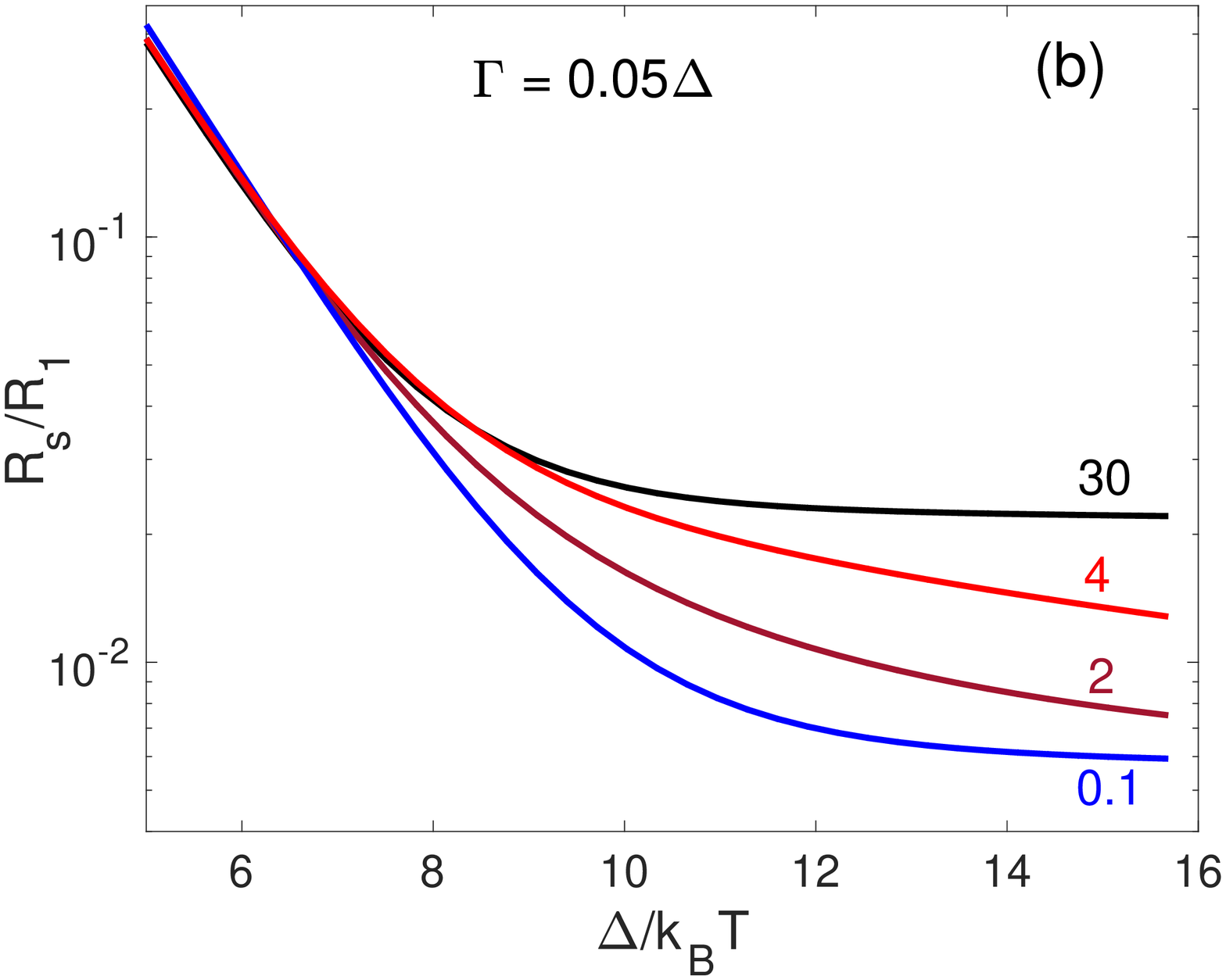}
   \caption{Arrhenius plots calculated from Eqs.~(\ref{eq:general_Rs_2})-(\ref{eq:IS}) for $\alpha=0.05$, $\lambda=4\xi_S$,  $\hbar\Omega= 11\Delta$, $D_n=D_s/3$,  $\beta=0.1,~2,~4,~30$, 
   and (a) $\Gamma=0.01\Delta$, (b) $\Gamma=0.05\Delta$. Here $R_1= \mu_0^2\omega^2\xi_S\lambda^2/2\rho_s$.}
   \label{fig10}
\end{figure}

Figures~\ref{fig10} show the Arrhenius plots of $\ln R_s(T)$ as functions of $\Delta/k_BT$ calculated from Eqs.~(\ref{eq:general_Rs_2})-(\ref{eq:IS}) for a thin N layer 
with $\alpha=0.05$, $\hbar\omega<\Gamma$ and different values of the interface barrier parameter $\beta$ varying from $\beta=0.1$ (weak resistive barrier) to $\beta=30$ (strong resistive barrier).
Behaviors of $R_s(T)$ for different values of $\Gamma=0.01$ (top panel) and $\Gamma=0.05$ (bottom panel) are qualitatively similar, so we 
focus on the evolution of $R_s(T)$ as a function of the control parameter $\beta$.  As $\beta$ increases, the surface resistance first decreases and then starts increasing with $\beta$ due to a subtle interplay of bulk and surface effects which will be discussed in the next subsection. However, at $\beta\gtrsim 1$, the surface resistance strongly increases with $\beta$, particularly at low temperatures $\Delta/k_BT > 10$, where $R_s(T,\beta)$ can increase by 1-2 orders of magnitude. The latter results from the RF dissipation in the N layer in which the proximity-induced superconductivity gets more and more suppressed with the increase of the interface resistive barrier.  At $\beta=4$, a noticeable change in the slope of $\ln R_s(T)$ around $\Delta/k_B\simeq 8-10$ results from switching from thermally-activated resistance controlled by the bulk gap $\Delta$ at high $T$ to the thermally-activated $R_s(T)$ controlled by the minigap $\epsilon_0$ in the thin N layer. As the temperature decreases further, the thermally-activated $R_s(T)$ approaches a residual resistance. 
Weakening the proximity-induced superconductivity in N layer at larger $\beta$ can significantly increase the residual resistance, as shown in Fig. \ref{fig10}.   

\subsection{Engineering the optimal density of states to minimize $R_s$}

\begin{figure}
\vspace{-2mm}	
   \includegraphics[width=7.3cm]{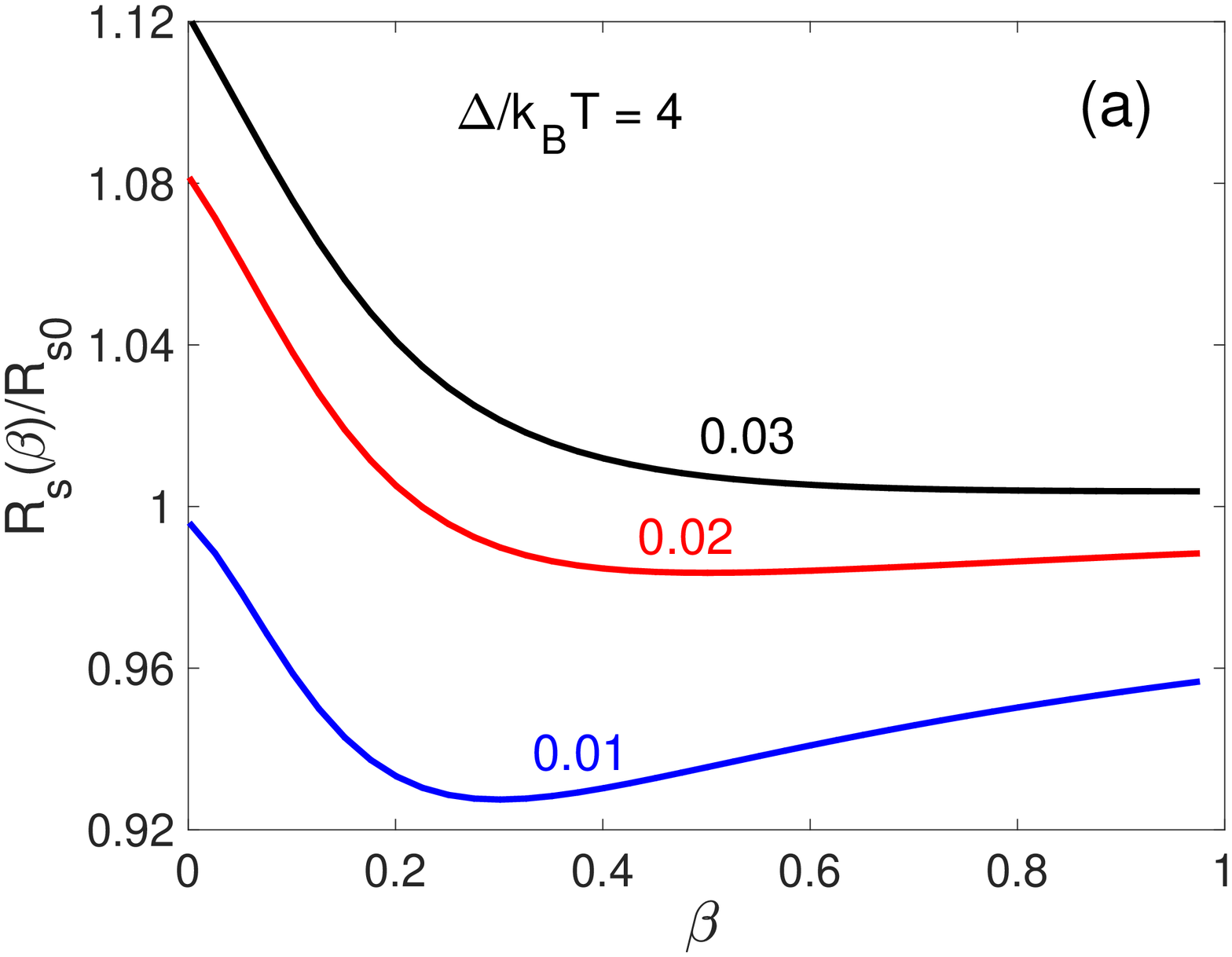}
   \vspace{-2mm}
   \includegraphics[width=7.3cm]{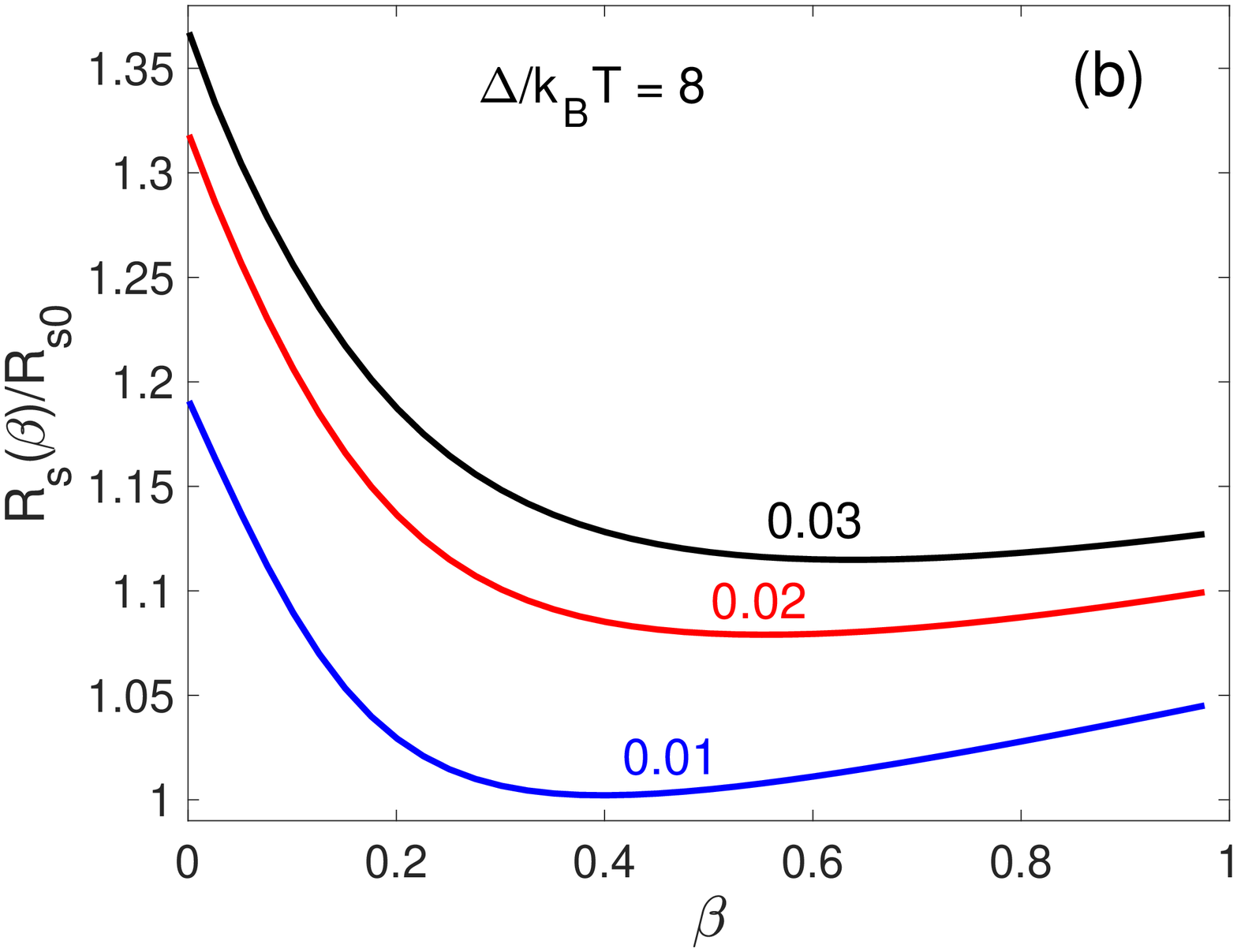}
   \vspace{-1mm}
   \includegraphics[width=7.3cm]{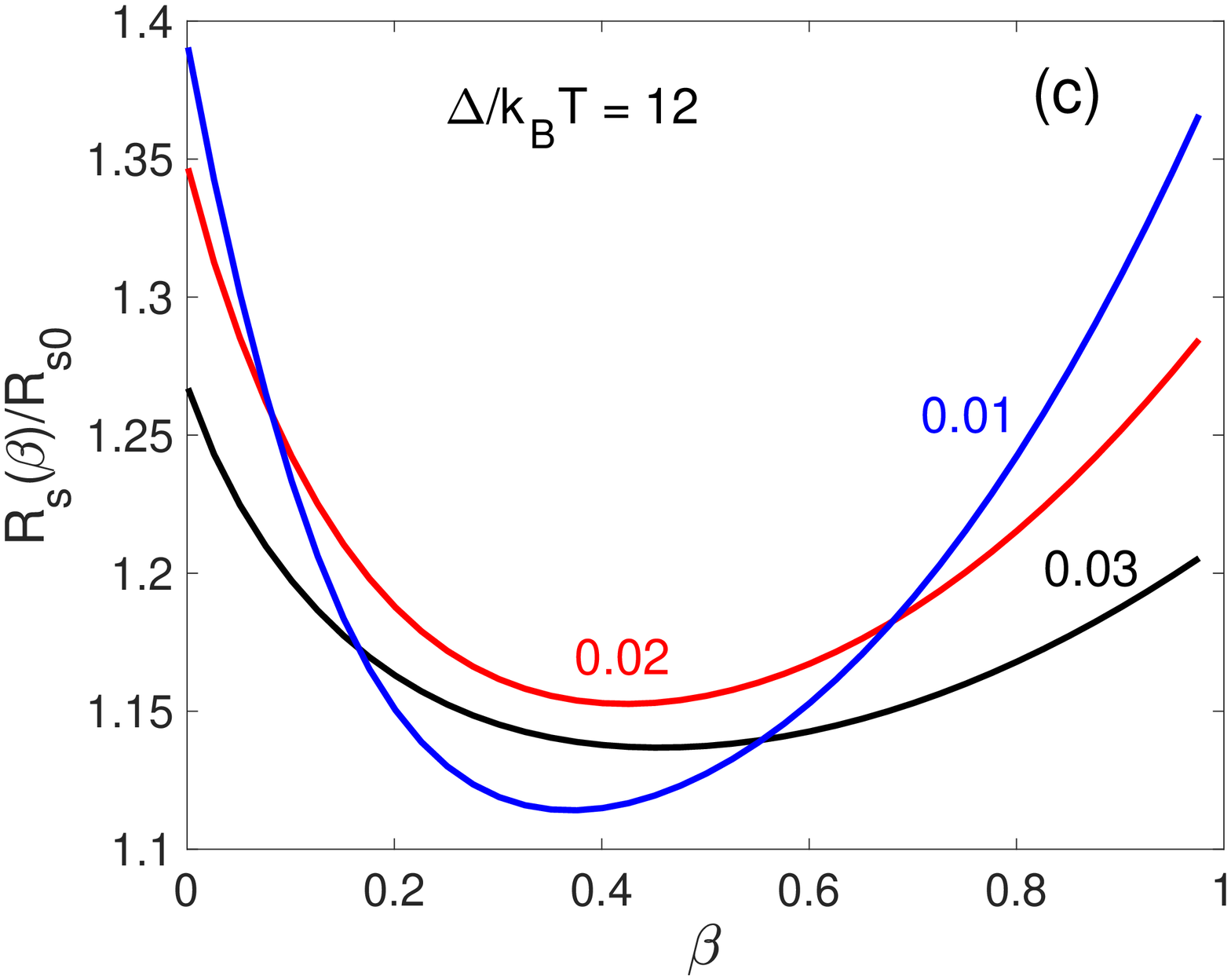}
   \vspace{-1mm}
   \includegraphics[width=7.3cm]{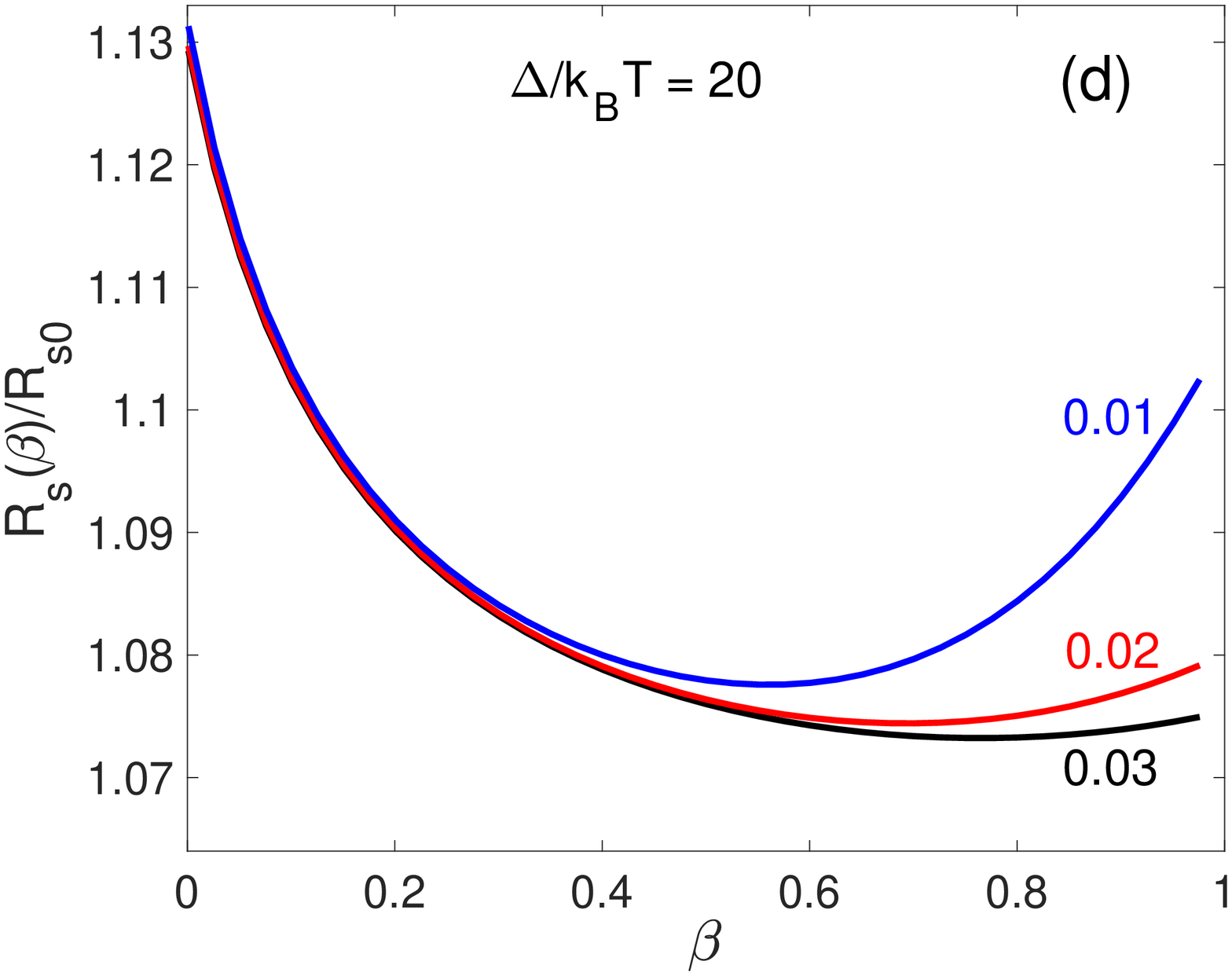}
   \vspace{-1mm}
   \caption{Minima in $R_s(\beta)$ as a function of the 
   interface barrier parameters $\beta$ calculated at $\Delta/k_BT=4,~8,~12,~20$, $\Gamma=0.01,~0.02, ~0.03$,  $\lambda=4\xi_S$,
   $D_n=0.1D_s$, $\alpha=0.05$, and $\hbar\Omega = 11\Delta$. Here $R_s(T,\beta,\Gamma)$ are normalized to their respective values $R_{s0}(T,\Gamma)$ for each $\Gamma$ in the absence of 
   N layer. }
   \label{fig11}
\end{figure}

The theory presented above suggests that DOS at the surface can be optimized to reduce $R_s$ by tuning the parameters of N layer. Shown in Fig. \ref{fig11} is an example of such optimization of $R_s(T,\Gamma,\beta)$ at different temperatures calculated for a thin dirty N layer with $D_n=0.1D_s$, $\alpha=0.05$, $\lambda=4\lambda_s$ and different values of $\Gamma/\Delta=0.01,~0,02,~0.03$.   Here all $R_s(\beta)$ curves are normalized to the respective values of $R_{s0}(T,\Gamma)$ for an ideal surface without N layer. The most noticeable features of these results is a minimum of $R_s(\beta)$ which shifts to larger $\beta$ as $T$ decreases.  Moreover, for the case of $\Delta/k_BT = 4$ shown in Fig. \ref{fig11}(a), the minimum value of $R_s$ at $\Gamma=0.01\Delta$ with N layer drops {\it below} the corresponding $R_{s0}$ for an ideal surface.      

\begin{figure}[tb]
   \includegraphics[width=9cm]{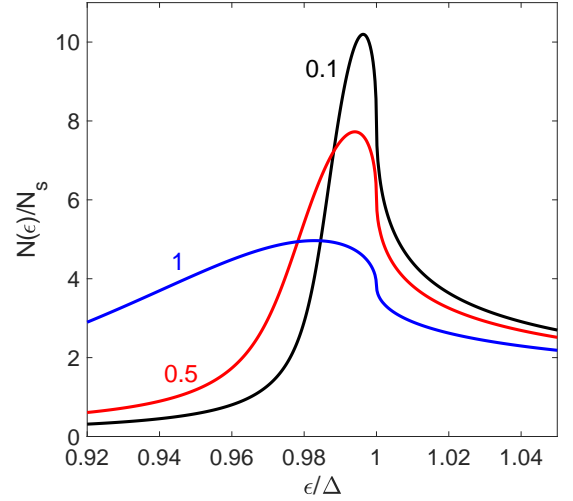}
   \caption{Behavior of $N(\epsilon)$ in a narrow energy range $\epsilon\approx \Delta$ at the S side of the NS interface calculated for $\beta=0.1,~0.5,~1$ at $\alpha=0.05$, $\hbar\Omega=11\Delta$, and $\Gamma=0$.  }
   \label{fig12}
   \end{figure}

The minimum in $R_s(\beta)$ mainly results from interplay of two effects. The first effect which causes $R_s$ to increase with $\beta$ is rather transparent: as the barrier parameter $\beta$ increases the proximity-induced superconductivity in N layer weakens, so the RF dissipation and $R_s$ increase. The second effect which causes the initial decrease of $R_s$ with $\beta$ results from the change in DOS around N layer.  As was pointed out in Refs. \onlinecite{ag_sust,agprl} and discussed in the previous subsections, a moderate broadening of the gap peaks in $N(\epsilon)$ due to either a finite quasiparticle lifetime $\hbar/\Gamma$ or magnetic impurities or current eliminates the BCS logarithmic divergence of $\sigma_1(\omega)$ at $\omega\to 0$ and reduces $R_s$. This mechanism also eliminates the BCS gap singularity in DOS around N layer even at $\Gamma=0$, as illustrated in Fig. 12 which shows that the peak in DOS broadens and decreases in amplitude as $\beta$ increases. At the same time, the N layer produces low-energy peaks in DOS at $\epsilon\approx\epsilon_0$, as shown in Fig. 4. It is not immediately clear if the reduction of $R_s$ due to the broadening of the DOS peaks at $\epsilon\approx \Delta$ would prevail over the increase of $R_s$ caused by the minigap peaks in DOS  at $\epsilon\approx \epsilon_0$.

Here we present a qualitative argument that the increase of the interface resistance $R_B$ not only produces a minimum in $R_s(\beta)$ but can reduce the overall surface resistance below $R_{s0}$ for an ideal surface. Indeed, the RF power in the proximity-coupled dirty $N$ layer which increase $R_s$ is proportional to the small thickness $d\ll \xi_S$ and also the conductivity $\sigma_n<\sigma_s$. The magnitude of low-energy tails in $N(\epsilon)$ in S region shown in Fig. 4b is also proportional to the small parameter $\alpha = d/\xi_S$. In turn, the decrease of $R_s(\beta)$ comes from the broadening of peaks in $N(\epsilon)$ at $\epsilon\approx\Delta$ in the bulk of S region. As follows from Eq. (\ref{dos_S}), the DOS disturbance $\delta n_S(x,\epsilon)$ produced by N layer extends into S region over the length $L=\xi_S[1-(\epsilon+i\Gamma)^2/\Delta^2]^{-1/4}$ which can be much larger than both $d$ and $\xi_S$. Indeed, $|L(\epsilon)|$ is maximum at $\epsilon=\sqrt{\Delta^2-\Gamma^2}$ for which
\begin{equation}
\!\delta n_s (x)\propto \exp\biggl[-\frac{x}{\xi_S^-}\biggl(\frac{\Gamma}{\Delta}\biggr)^{1/4}\biggr]\!\cos\biggl[\frac{x}{\xi_S^+}\biggl(\frac{\Gamma}{\Delta}\biggr)^{1/4}\!\!\!\!+\varphi \biggr],
\label{fig12}
\end{equation}
where $\xi_S^\pm=\xi_S[2(\sqrt{2}\pm 1)]^{1/2}$, and $\varphi$ is a phase shift. Hence, the broadening effect responsible for the decrease of $R_s$ is produced by a long-range disturbance in DOS in S region where the decay  length $L\sim\xi_S(\Delta/\Gamma)^{1/4}$ increases as $\Gamma$ decreases. This bulk contribution coming from a layer of width $\sim L \gg d$ in S region exceeds the short-range contributions from N layer, so that a thin dirty N layer on the surface of superconductor with $\Gamma\ll \Delta$ can reduce its surface resistance. As $\Gamma$ increases, $\delta n_s(x,\Gamma)$ becomes more short-range, and $R_s(\beta)$ eventually exceeds $R_{s0}$ at all $\beta$, in agreement with Fig. \ref{fig11}.  This behavior is a manifestation of a counterintuitive reduction of $R_s$ by pairbreaking mechanisms which normally reduce $T_c$ but broaden the gap peaks in $N(\epsilon)$ ~ \cite{ag_sust}. This effect demonstrated here for the  Dynes model and paramagnetic impurities (see Figs. \ref{ri} and \ref{rsmagg}) can also cause a microwave reduction of surface resistance, namely a decrease of $R_s(H)$ with the amplitude of the RF field~ \cite{agprl}.  

As $T$ decreases, the minima in $R_s(\beta)$ shown in Fig. \ref{fig11} shift to larger $\beta$, so the optimal value of the interface boundary resistance $R_B$ at which $R_s(\beta)$ is minimum is different at different temperatures.  For the calculations presented in Fig. \ref{fig10}, the minimum $R_s(\beta)$ can drop by $\simeq 3-15\%$ relative to $R_{s0}$, depending on the particular values of $T$ and $\Gamma$.  At $\beta \gtrsim 1$ the surface resistance increases strongly with $\beta$ and exceeds $R_{s0}$ by several orders of magnitude at $\beta\gg 1$ (see Fig. \ref{fig10}). Thus, optimization of $R_B$ by different materials treatments can be really important to reduce the surface resistance.     

\vspace{-2mm}
\section{Discussion}

Our results show that a non-ideal surface can locally broaden the gap peaks in DOS, resulting in a a non-exponential temperature dependence of the surface impedance $Z(T)$ at $T\ll T_c$. Because the main broadening effect can occur in a layer much thinner than the field penetration depth, tunneling surface 
probes do not really give all information about the features of DOS on the relevant scales of the London penetration depth which determine $Z(T)$. For instance Fig. 2 shows that 
the broadening effect caused by reduction of the pairing constant at the surface can be much stronger than in the bulk where $N(\epsilon)$ has much sharper peaks. 
The broadening of DOS at the surface can become much more pronounced if a thin proximity-coupled normal layer is present  (see Figs. 4-6). Thus, fitting the tunneling data with 
Eq. (2) and extracting the "effective" $\Gamma$ to describe the low-$T$ surface impedance can be misleading, but a combination of $Z(T)$ and tunneling 
measurements in a sufficiently broad temperature range may offer a possibility to separate the surface and bulk contributions to $Z(T)$.  

Measurements of $Z(T)$ in a broad range of temperatures is really important for getting the correct physical picture, as opposed to fitting the experimental data with the phenomenological Eq. (1) in a limited temperature window and  extrapolating the results to lower $T$. For instance, the behaviors of $R_s(T)$ in Fig. \ref{fig10} (b) at $4<\Delta/k_BT<9$ for $\beta=4$ and $\beta=30$ are nearly the same, so using Eq. (1) would suggest the residual resistance to be close to $R_i$ at $\beta=30$ in both cases. However, the actual temperature dependencies of $R_s(T)$ at $\Delta/k_BT > 9$ are markedly different:  $R_s(T)$ at $\beta=30$ reaches the residual resistance plateau at $\Delta/k_BT\simeq 10$, whereas $R_s(T)$ for $\beta=4$ keeps decreasing exponentially with the Arrhenius slope controlled by  a smaller minigap $\epsilon_0$ in the N layer, so that the residual plateau is reached at much lower temperatures $\Delta /k_BT > 20$.  

The "two-exponential" temperature dependence of $R_s(T)$ controlled by the bulk and the surface gaps $\Delta$ and $\epsilon_0$ can mimic the low-temperature behavior of $Z(T)$ in multi-band superconductors with different gaps residing on different sheets of the Fermi surface, as characteristic of MgB$_2$ \cite{mgb2a,mgb2b,mgb2c} or iron pnictides \cite{fbs1,fbs2,fbs3,fbs4}. Because deviations from the single-band s-wave exponential temperature dependence of the London penetration depth $\lambda(T)$ has been often regarded as evidence of unconventional pairing symmetry, be it the $d$-wave pairing in cuprates \cite{lam_d} or $s_\pm$ pairing in pnictides \cite{tb1,tb2}, the surface contribution and bulk subgap states may complicate an analysis of experimental data. Indeed, Eq. (\ref{lam}) shows that a thin N layer on the oxidized surface of a conventional single-band s-wave superconductor can result in a "two-gap" temperature dependence of $\lambda(T)$, while the subgap states in the phenomenological Dynes model can result in a quadratic temperature dependence (\ref{lam1}), similar to that has been observed on pnictides where it was associated with subgap impurity states for multiband pairing \cite{tb1,tb2}.  Surface non-stoichiometry and interface strains in true multiband superconductors can further complicate extracting the gaps and revealing pairing symmetries. Yet the surface contribution to $\lambda$ coming from the disturbance of superfluid density in a thin N layer is generally much smaller than its contribution to $R_s$ determined by long-range tails of $N(\epsilon,x)$ in the bulk at $\epsilon\approx \Delta$.       

Although the minigap $\epsilon_0$ in Eq. (\ref{mgap}) for a weakly coupled N layer is independent of superconducting parameters, $\epsilon_0(T)$ can depend on $T$ even at $T\ll T_c$ if $R_B(T)$ changes with $T$, resulting  in a non-exponential temperature dependencies of $\lambda(T)$ and $R_s(T)$.  It is well-known that the interface resistance $R_B(T)$ can either increase or decrease with temperature, depending on the materials heat treatment which can change $R_B$ by several orders of magnitude, as was, for example, shown for the YBCO-Ag interface~ \cite{int1,int2}. The complex physics and materials science of the Schottky barrier between different materials is not well understood~ \cite{barrier}, but the essential dependence of $R_s(T)$ on the interface resistance could be used to optimize $R_s$ by tuning the properties of N layer.         

This work shows that a non-ideal surface can significantly contribute to the residual surface resistance, which becomes an integral part of the surface resistance taking into account realistic broadening of the DOS peaks. Clearly, a thin pairbreaking layer or a weakly-coupled normal layer at the surface can radically (by orders of magnitude) increase $R_i$ as compared to an ideal surface with only bulk broadening mechanisms. However, $R_s(T)$ can be reduced by optimizing DOS at the surface by tuning the properties of a proximity-coupled N layer at the surface. For weak bulk broadening $\Gamma\ll \Delta$, the results shown in Fig. \ref{fig11} suggest that a thin N layer can surprisingly reduce $R_s$ as compared to an ideal surface, if the interface contact resistance $R_B$ is within a sweet spot of the parameters for which $\beta\simeq 0.2-1$. Yet this theory also shows that there is no universal value of $R_B$ which provides optimal $R_s$ for all $T$: an optimal $R_B$ for one temperature may not be as good for another.  

To estimate $R_B$ corresponding to $\beta=1$ in Nb, we assume $N_n=N_s$ and present Eq. (\ref{beta}) in the form, $\beta = 16 R_B d/R_K\lambda_F^2\xi_0$, where  $R_K=h/e^2=26$ k$\Omega$, 
$\lambda_F=h/m^*v_F$ is the Fermi wavelength, $\xi_0=\hbar v_F/\pi\Delta$ is the coherence length in the clean limit.  Taking $\xi_0=40$ nm, $d=1$ nm, $\lambda_F=5.3\cdot 10 ^{-10}$ m~ \cite{ashkroft}, we obtain that $\beta\lesssim 1$ corresponds to $R_B\lesssim  R_K\lambda_F^2\xi_0/16 d\simeq 1.8\cdot 10^{-14}$ $\Omega\cdot$m$^2$, about one-two orders of magnitude smaller than the lowest contact resistance of the YBCO/Ag interfaces, $R_B \sim 10^{-13}-10^{-12}$ $\Omega\cdot$m$^2$ ~\cite{int2}. 

Tuning $R_s(T)$ by changing the properties of the surface N layer is based on the fact that the idealized DOS with $\Gamma=0$ does not produce the lowest $R_s$ because the BCS conductivity $\sigma_1(\omega)$ diverges logarithmically at $\omega\to 0$. Thus, $R_s$ can be lowered by pairbreaking mechanisms which suppress $T_c$ but broaden the gap peaks in DOS~ \cite{ag_sust}, for example, by incorporating a small concentration of paramagnetic impurities (see Fig. \ref{rsmagg}). Currently, the physics of subgap states in the bulk is not well understood, so the materials means of tuning the Dynes parameter $\Gamma$ are unclear. However,  tuning the thickness and conductivity of the surface N layer and the interface resistance $R_B$ by different materials treatments, impurity managements and surface nano-structuring appears more technologically viable. Our calculations show that $R_s$ can indeed by optimized by tuning the interface resistance to provide $\beta\simeq 0.2-1$. These results may help understand the effect of  low-temperature heat treatment on the reduction of $R_s$ observed on the Nb resonant cavities \cite{clare,padams} and suggest ways of reducing the RF dissipation in thin film superconducting structures and micro resonators.      

\vspace{-5mm}
\appendix

\section{Derivation of Eq.~(\ref{eq:ex_deltaDelta_sol})} \label{appendix:simple_ex}

Equation (\ref{eq:ex_thermodynamic_Usadel}) is solved by the Fourier transform $\delta\theta(x)=\sum_k\delta\theta_k\cos kx$ and $\delta \Delta(x)=\sum_k\delta\Delta_k\cos kx$ which automatically satisfies $\delta\theta'(0)=0$:
\begin{equation}
\delta\theta_k = \frac{\omega_n \delta\Delta_k}{k_{\omega}^2 (k_{\omega}^2 + k^2)},
\label{aeq:ex_deltatheta_k}
\end{equation}
where $k_\omega=(1+\omega_n^2)^{1/4}$. The Fourier transform of Eq.~(\ref{eq:ex_self_consistency}) yields:
\begin{equation}
\delta \Delta_k = 2\pi T g \sum_{\omega_n>0}^{\Omega} \cos\theta_{\infty} \delta\theta_k + 2\pi T \delta g_k \sum_{\omega_n>0}^{\Omega} \sin\theta_{\infty} .
\label{aeq:ex_deltaDelta_k_0}
\end{equation}
Substituting Eq.~(\ref{aeq:ex_deltatheta_k}) into Eq.~(\ref{aeq:ex_deltaDelta_k_0}) and using the gap equation $1=2\pi T g_{\infty} \sum_n \sin\theta_{\infty}$, 
we obtain 
\begin{gather}
\delta \Delta_k = \frac{\delta g_k}{g^2 S(k)}, 
\label{aeq:ex_deltaDelta_k} \\
S(k) = 2\pi T \sum_{\omega_n>0}^{\infty} \frac{k^2\sqrt{\omega_n^2+1}+1}{(\omega_n^2+1)(k^2+\sqrt{\omega_n^2+1})}.
\label{aeq:S_q}
\end{gather}
At $k_BT \ll \Delta$, the summation in Eq. (\ref{aeq:S_q}) can be replaced with integration: 
\begin{gather}
S(k) 
= \int_0^{\infty}\frac{(k^2\sqrt{\omega^2+1}+1)d\omega}{(\omega^2+1)(k^2+\sqrt{\omega^2+1})}= \nonumber \\
\begin{cases}
\frac{\pi}{2k^2}-\frac{2}{k^2}\sqrt{1-k^4} \tan^{-1} \sqrt{\frac{1-k^2}{1+k^2}} & (k^2<1) \\
\frac{\pi}{2k^2} + \frac{1}{k^2}\sqrt{k^4-1} \ln(k^2+\sqrt{k^4-1}) & (k^2>1).
\end{cases}
\label{aeq:S_q_int}
\end{gather}

\section{Derivation of Eqs.~(\ref{eq:reducedBCS_solution}) and (\ref{eq:reducedBCS_theta0})} \label{appendix:reducedBCS_retarded}

Consider the Usadel equation
\begin{equation}
\theta'' = i[1 -\Psi \delta(x)]\cosh\theta -  i \epsilon \sinh\theta  
\label{aeq:reducedBCS_real_Usadel_1}
\end{equation}
with the boundary conditions $\theta(+\infty)=\theta_\infty$ and $\theta'(\infty)=0$. Another boundary condition at $x=+0$ is obtained by 
integrating Eq.~(\ref{aeq:reducedBCS_real_Usadel_1}) from $0$ to $+0$:
\begin{equation}
\theta'(+0) = -i\Psi \cosh\theta_0 , \label{aeq:new_BC_reducedBCS}
\end{equation} 
Multiplying both sides of Eq.~(\ref{aeq:reducedBCS_real_Usadel_1}) by $\theta'$ and integrating it from $x=+0$ to $\infty$, 
we obtain: 
\begin{equation}
\frac{\theta'^2}{2}=i\sinh\theta -i\epsilon\cosh\theta -i\sinh\theta_\infty +i\epsilon\cosh\theta_\infty,
\label{temp}
\end{equation}
where the last two terms in the right hand side provide the boundary condition $\theta'(\infty)=0$. Dividing both sides of Eq. (\ref{temp}) by $k_\epsilon^2=\sqrt{1-\epsilon^2}$ and using there $\sinh\theta_\infty=1/\sqrt{\epsilon^2-1}$ and $\cosh\theta_\infty =\epsilon/\sqrt{\epsilon^2-1}$, Eq. (\ref{temp}) can be reduced to $\theta'^2=2k_\epsilon^2[\cosh(\theta-\theta_\infty)-1]$. Hence,
\begin{eqnarray}
\theta' = -2 k_{\epsilon} \sinh \frac{\theta -\theta_{\infty}}{2}, \label{aeq:reducedBCS_theta'}
\end{eqnarray}
where the minus sign was taken to provide the solution which approaches $\theta_\infty$ at $x\to\infty$.
The self-consistency equation for $\theta_0$ is obtained by taking the limit $x\to+0$ in Eq. (\ref{aeq:reducedBCS_theta'}) and expressing $\theta'(+0)$ using Eq.~(\ref{aeq:new_BC_reducedBCS}):
\begin{eqnarray}
i\Psi \cosh\theta_0  = 2 k_{\epsilon} \sinh \frac{\theta_0 -\theta_{\infty}}{2}. 
\end{eqnarray}
Integration of Eq.~(\ref{aeq:reducedBCS_theta'}) with the boundary condition $\theta(+0)=\theta_0$ yields $\theta(x)$ in the form:   
\begin{eqnarray}
\tanh\frac{\theta(x)-\theta_{\infty}}{4} = \tanh\left[ \frac{\theta_0-\theta_{\infty}}{4}\right]  e^{-k_{\epsilon}x}. 
\end{eqnarray}
%

\section{Derivation of Eqs.~(\ref{eq:NS_solution}) and (\ref{eq:NS_theta0})} \label{appendix:NS_im_Usadel}

The thermodynamic Usadel equation
\begin{eqnarray}
\theta'' = -[1- \Psi \delta(x+0)] \cos \theta + \omega_n\sin\theta 
\label{aeq:NS_Im_Usadel_1}
\end{eqnarray}
is solved in the same way as in Appendix~\ref{appendix:reducedBCS_retarded}. Integration of Eq.~(\ref{aeq:NS_Im_Usadel_1}) from $0$ to $+0$ gives:
\begin{eqnarray}
\theta'(+0) = \omega_n\Phi \sin\theta_0  + \Psi \cos \theta_0. 
\label{aeq:BC_NS2}
\end{eqnarray}
Next, we multiply both sides of Eq.  (\ref{aeq:NS_Im_Usadel_1}) by $\theta'$ and integrate it from $x>0$ to $\infty$ with the 
boundary conditions $\theta(\infty)=\theta_\infty$ and $\theta'(\infty)=0$: 
 \begin{equation}
\frac{\theta'^2}{2}=-\sin\theta -\omega_n\cos\theta +\sin\theta_\infty +\omega_n\cos\theta_\infty.
\label{temp1}
\end{equation}
Dividing both sides of Eq. (\ref{temp1}) by $k_\omega^2=\sqrt{1+\omega_n^2}$ and using $\cos\theta_\infty=\omega_n/k_\omega^2$, $\sin\theta_\infty = 1/k_\omega^2$, yields:
\begin{equation}
\theta'(x) = -2 k_{\omega} \sin \frac{\theta(x) -\theta_{\infty}}{2} . \label{aeq:NS_theta'}
\end{equation}
The self-consistency equation for $\theta_0$ is obtained by taking the limit of $x\to +0$ in Eq. (\ref{aeq:NS_theta'}) 
and expressing $\theta'(+0)$ in terms of $\theta_0$ using Eq.~(\ref{aeq:BC_NS2}):
\begin{equation}
\omega_n \Phi \sin\theta_0  + \Psi \cos \theta_0  + 2 k_{\omega} \sin \frac{\theta_0 -\theta_{\infty}}{2} =0 , 
\end{equation}
Integration of Eq. (\ref{aeq:NS_theta'}) with the boundary condition $\theta(+0)=\theta_0$ then yields $\theta(x)$ in the form:
\begin{equation}
\tan\frac{\theta(x)-\theta_{\infty}}{4} = \tan \biggl[\frac{\theta_0-\theta_{\infty}}{4}\biggr]  e^{-k_{\omega}x} . 
\end{equation}
%

\section{Evaluation of $\Psi(T)$} \label{appendix:SN_SD}
We evaluate $\Psi$ and $\Phi$ for a thin N layer, for which $\alpha\ll 1$ and the sums for $S_D$ and $S_N$ in Eq. (\ref{eq:self_consistency_gamma2}) converge at  
$\omega_n \gg 1$ so that $k_\omega=(1+\omega_n^2)^{1/4}\to \sqrt{\omega_n}$.  However, at $\omega_n\gg 1$ which give the main contribution to the sums (\ref{eq:SN}) and (\ref{eq:SD}), 
we have $\cos\theta_0\to\cos\theta_\infty\to 1$, and  
\begin{equation}
\Phi\simeq \frac{\alpha}{1+\beta\omega_n}, \qquad \omega_n\gg 1.
\label{Ph}
\end{equation}
Using Eq. (\ref{Ph}), the factor $k_\omega^3+\omega_n^2\Phi$ in the denominators of $S_N$ and $S_D$ at $\omega_n\gg 1$ can be written in the form:
\begin{equation}
k_\omega^3+\omega_n^2\Phi \to \omega_n^{3/2}\left(1+\frac{\alpha\sqrt{\omega_n}}{1+\beta\omega_n}\right)
\label{fact}
\end{equation}  
If $\alpha\ll 1$ and $\beta>1$, the second term in the parenthesis is negligible at any $\omega_n$, but it may become essential at $\beta\ll 1$ and $\omega_n\gg 1$. 
This term cannot exceed $\alpha/2\sqrt{\beta}$ so $\omega_n^2\Phi$ in Eq. (\ref{fact}) can be neglected if:
\begin{equation}
\beta > \alpha^2/4
\label{condit}
\end{equation}
In the case of $\alpha\ll 1$ considered in this work, the condition (\ref{condit}) is satisfied in a wide range of the parameters comprising both limits of strongly coupled 
$(\beta\ll 1)$ and weakly coupled $(\beta\gg 1)$ N layer. If Eq. (\ref{condit}) holds, we can neglect $\omega_n^2\Phi$ in the denominator of Eq. (\ref{eq:SD}), so that
\begin{gather}
1-S_D=1-2\pi Tg\sum_{\omega_n>0}^\Omega \frac{\omega_n^2}{(1+\omega_n^2)^{3/2}} = 
\nonumber \\
1-2\pi T \sum_{\omega>0}^\Omega \frac{g}{\sqrt{1+\omega_n^2}}+2\pi T \sum_{\omega_n>0}^\Omega \frac{g}{(1+\omega_n^2)^{3/2}}.
\label{sum}
\end{gather}
The first two terms in the last line of Eq. (\ref{sum}) cancel out as they represent the BCS gap equation, thus:
\begin{equation}
1-S_D=2\pi T\sum_{\omega_n>0}^\infty \frac{g}{(1+\omega_n^2)^{3/2}}.
\label{sumS}
\end{equation} 
Substitution of Eqs. (\ref{eq:Phi_def}) and (\ref{sumS}) into Eq. (\ref{eq:self_consistency_gamma2}) yields:
\begin{gather}
\!\!\Psi=\sum_{\omega>0}^\Omega \frac{2\pi T\alpha\omega_n^2}{I_D(1+\omega_n^2)^{3/2}\sqrt{1+\beta^2\omega_n^2+2\beta\omega_n\cos\theta_\infty}},
\label{Ps}\\
I_D=2\pi T\sum_{\omega_n>0}(1+\omega_n^2)^{-3/2},
\label{I}
\end{gather}
where $g$ cancels out. Here we set $\cos\theta_0\to\cos\theta_\infty=\omega_n/\sqrt{1+\omega_n^2}$, which is a good approximation 
for a thin N layer, as was shown in Sect. III B. Equations (\ref{Ps}) and (\ref{I}) combined with the bulk BCS gap equation determine the temperature dependence of $\Psi(T)$.
At $T\ll T_c$ the summation in Eqs. (\ref{Ps}) and (\ref{I}) can be replaced with integration, giving $I=1$ and:  
\begin{equation}
\Psi=\int_0^\Omega \frac{\alpha\omega^2d\omega}{(1+\omega^2)^{3/2}\sqrt{1+\beta^2\omega^2+2\beta\omega\cos\theta_\infty}}
\label{Ps0}
\end{equation}
If $\beta\gg 1$, the square root in Eq. (\ref{Ps0}) becomes $\beta\omega$, and 
\begin{equation}
\Psi=\alpha/\beta, \qquad \beta\gg 1.
\label{Ps1}
\end{equation}   
If $\beta < 1$ the main contribution to the integral comes from the region $\omega > 1$ where $\cos\theta_\infty\to 1$, so that:
\begin{equation}
\Psi=\int_0^\Omega \frac{\alpha\omega^2d\omega}{(1+\omega^2)^{3/2}(1+\beta\omega)}, \qquad \beta \ll 1
\label{Ps2}
\end{equation}
Calculation of this integral yields Eq. (\ref{Psi}) which also comprises the large-$\beta$ limit (\ref{Ps1}) and gives a good approximation of $\Psi$ at $\beta\simeq 1$. Temperature corrections to 
Eq. (\ref{Psi}) are exponentially small at $T<T_c/2$. 

For a high-transparency NS boundary, $\beta\ll \alpha^2/4$, we can set $\beta=0$ and obtain:
\begin{gather}
\Psi=\frac{\alpha g S}{1-g S}, \qquad \beta\ll \alpha^2/4,
\label{p1} \\
S=2\pi T\sum_{\omega_n>0}^\Omega\frac{\omega_n^2}{(1+\omega_n^2)^{3/4}[(1+\omega_n^2)^{3/4}+\alpha\omega_n^2]}.
\label{s}
\end{gather}
If $\alpha \ll (\Delta/\hbar\Omega)^{1/2}\ll 1$, the term $\alpha\omega_n^2$ in the denominator of $S$ can be neglected. Replacing the sum with the integral and using the BCS gap equation yields:
\begin{gather}
S=\int_0^\Omega\frac{\omega^2d\omega}{(\omega^2+1)^{3/2}}=\frac{1}{g}-1.
\label{so} \\
\Psi=\alpha\left(\frac{1}{g}-1 \right), \qquad \alpha\ll (\Delta/\hbar\Omega)^{1/2},
\label{p2}
\end{gather}    
which reduces to Eq. (\ref{psss}). If $(\Delta/\hbar\Omega)^{1/2}\ll\alpha\ll 1$, the convergence of the sum in Eq. (\ref{s}) is provided by $\alpha\omega_n^2$ in the denominator.
Here we introduce an intermediate cutoff $\omega_c$ such as $1\ll \omega_c \ll \alpha^{-2}$, set $\Omega\to\infty$, and split $S$ in Eq. (\ref{s}) into two parts:
\begin{gather}
S=\int_0^{\omega_c}\frac{\omega^2 d\omega}{(1+\omega^2)^{3/2}}+\int_{\omega_c}^\infty\frac{d\omega}{\omega(1+\alpha\sqrt{\omega})} =
\nonumber \\
-\frac{\omega_c}{\sqrt{1+\omega_c^2}}+\sinh^{-1}\omega_c+2\ln\frac{1+\alpha\sqrt{\omega_c}}{\alpha\sqrt{\omega_c}}.
\label{sint}
\end{gather}
In the limit of $\alpha\sqrt{\omega_c}\ll 1$ and $\omega_c\gg 1$ the auxiliary parameter $\omega_c$ in Eq. (\ref{sint}) cancels out, giving
\begin{equation}
S\simeq \ln(2/\alpha^2)-1.
\label{sss} 
\end{equation} 
For Nb, the parameter $(\Delta/\hbar\Omega)^{1/2}\approx 0.3$ is not particularly small, 
so the condition under which Eq. (\ref{p2}) is valid overlaps with the condition of applicability of the thin-N layer approximation of this work $\alpha\ll 1$. 
  
\section{Complex conductivity.} \label{appendix:sigma}

Complex conductivity $\sigma(\omega)=\sigma_1(\omega)-i\sigma_2(\omega)$ of a dirty superconductor in the local London limit can be expressed in terms of retarded Green's functions \cite{nam,kopnin,proxi2}:
\begin{gather}
\sigma_1=\frac{\sigma_s}{\hbar\omega}\int_{-\infty}^\infty  d\epsilon \bigl[f(\epsilon)-f(\epsilon+\hbar\omega)\bigr]\times \nonumber \\ 
\!\!\!\![{\rm Re}G^R(\epsilon){\rm Re}G^R(\epsilon+\hbar\omega)+{\rm Re}F^R(\epsilon){\rm Re}F^R(\epsilon+\hbar\omega)],  
\label{sig1} \\
\sigma_2=\frac{\sigma_s}{\hbar\omega}\int_{-\infty}^\infty d\epsilon \tanh\frac{\epsilon}{2k_BT}\times \nonumber \\
\!\!\!\![{\rm Re}G^R(\epsilon){\rm Im}G^R(\epsilon+\hbar\omega)+{\rm Re}F^R(\epsilon){\rm Im}F^R(\epsilon+\hbar\omega)],
\label{sig2}
\end{gather}
where $f(\epsilon)$ is the Fermi distribution function. The surface 
impedance in the local limit $\lambda\gg\xi$ is then: 
\begin{equation}
Z=i\omega\mu_0\lambda + R_s=\left[\frac{i\mu_0\omega}{\sigma_1-i\sigma_2}\right]^{1/2}.
\label{zin}
\end{equation}    
At $k_BT\ll\Delta$ and $\hbar\omega\ll \Delta$, we have $\sigma_1\ll \sigma_2$, so Eq. (\ref{zin}) defines the London penetration depth 
$\lambda=(\mu_0\omega\sigma_2)^{-1/2}$.   

A quasi-static $\lambda$ can be obtained from the supercurrent density $\textbf{J}$ in the Matsubara representation:  
\begin{equation}
\textbf{J}=-\frac{4\pi k_BT}{\hbar\rho_s} \textbf{Q} \sum_{\omega_n>0}|F(\omega_n)|^2, 
\label{J}
\end{equation}
where $\textbf{Q}=\textbf{A}+(\phi_0/2\pi)\nabla\chi$, and $\chi$ is the phase of the order parameter.   
Since $\textbf{J}=-\textbf{Q}/\mu_0\lambda^2$, we have: 
\begin{equation}
\frac{1}{\lambda^2}=\frac{4\pi\mu_0 k_BT}{\hbar\rho_s} \sum_{\omega_n>0}|F(\omega_n)|^2.
\label{LL}
\end{equation}
Substituting here $|F|^2=\Delta^2/(\Delta^2+\hbar^2\omega_n^2)$ at $\Gamma=0$ yields Eq. (\ref{lam0})
which also follows from Eq. (\ref{sig2}) in the limit of $0<\hbar\omega\ll\Delta$ and $\Gamma=0$ where the integrand is nonzero at $-\Delta-\hbar\omega < \epsilon < -\Delta$. Hence   
$\sigma_2=(\pi\sigma_s\Delta/\hbar\omega)\tanh(\Delta/2k_BT)$, so that $\lambda^{-2}=\mu_0\omega\sigma_2$  reduces to Eq. (\ref{lam0}).

\vspace{-3mm}
\begin{acknowledgments}
The work of A.G. was supported by NSF under Grant PHY-1416051.
The work of T.K. was supported by Japan society for the promotion of science (JSPS) Grant-in-Aid for Young Scientists (A) $\# {\rm 17H04839}$ and 
JSPS Grant-in-Aid for Challenging Exploratory Research, $\# 26600142$.  

\end{acknowledgments}

\end{document}